# Comparative Analysis of Distributed and Parallel File Systems' Internal Techniques

Viacheslav Dubeyko

# Content

**TERMINOLOGY AND ABBREVIATIONS**

| Terminology | Description |
|---|---|
| [GPFS] | The General Parallel File System (GPFS) is a high-performance shared-disk clustered file system developed by IBM. |
| [GFS2] | Redhat Global File System (GFS2) is a shared disk file system for Linux computer clusters. |
| [CXFS] | The CXFS file system (Clustered XFS) is a proprietary shared disk file system designed by Silicon Graphics (SGI) specifically to be used in a Storage area network (SAN) environment. |
| [Lustre] | Lustre is a parallel distributed file system, generally used for large scale cluster computing. The name Lustre is a portmanteau word derived from Linux and cluster. |
| [PVFS] | The Parallel Virtual File System (PVFS) is an open source parallel file system. |
| [Hadoop] | Hadoop provides a distributed file system and a framework for the analysis and transformation of very large data sets using the MapReduce paradigm. |
| [MapReduce] | A distributed data processing model and execution environment that runs on large clusters of commodity machines. |
| [HDFS] | A distributed filesystem that runs on large clusters of commodity machines. |
| [InterMezzo] | InterMezzo is obsolete distributed file system with a focus on high availability. |
| [PlasmaFS] | PlasmaFS is a distributed filesystem for large files, implemented in user space. |
| [Ceph] | Ceph is a distributed file system that architecture is based on the assumption that systems at the petabyte scale are inherently dynamic. |
| [OCFS2] | OCFS2 (The Oracle Clustered File System) is a general-purpose shared-disk cluster file system for Linux capable of providing both high performance and high availability. |
| [Google File System] | GFS (Google File System) is designed as a distributed file system to be run on clusters up to thousands of machines. |

# 1 Introduction

The objective of the document is to compare different types of distributed file systems' features and internal techniques with the purpose of achieving understanding of advantages and disadvantages these techniques for the data mining use-case. The comparison results can be an analytical basis for a new distributed file system's design elaboration.

# 2 Comparative Analysis Methodology

The comparative analysis can be divided on several phases:

1. **File system's features classification**. This phase includes division of features between different classes on the basis of functional difference.

2. **Compare features of one class for the case of different file systems**. This phase includes efforts of confrontation different file systems' techniques of one class (for example, reliability). The purpose of the phase is to elaborate vision of possible ideological alternatives for this feature class, potential use cases of feature using, advantages and disadvantages of such solution.

3. **Data mining use-case analysis**. This phase includes efforts of distinguishing the key specifics and peculiarities of data mining algorithms from the distributed file systems point of view. The goal of the phase is to make dichotomy of internal nature of such class of algorithms with elaboration of vision of requirements to distributed file system's internal techniques.

4. **Select promising features**. This phase includes choosing promising features for a new file system architecture with analysis how it can be used with maximum efficiency for the data mining use-case.

# 3 File System Features Classification

Every file system includes technical approaches and architectural decisions which to make the essence of its internal techniques and to define the efficiency of file system's operations under concrete workload. Such technical approaches and architectural decisions can be identified as file system's features. A file system's feature has purpose to achieve some operational efficiency or to provide any services to end-user. Thereby, it is possible to classify the file system features from the point of view of concrete feature's goal.

It is possible to distinguish such the most important feature classes:

1. **Architectural features**. These features characterize a file system's design and describe the vision of principal architectural approaches that to define a file system's components and essence of internal interactions between of its.

2. **Performance features**. These features characterize a file system's internal techniques that are used for achieving high performance.

3. **Reliability features**. Reliability is the ability of a system or component to perform its required functions under stated conditions for a specified period of time. Reliability features are internal techniques and approaches that make possible to oppose against unfavourable factors (for example, Sudden Power-Off) and to keep data in consistent state.

4. **High-availability features**. High availability is a system design approach and associated service implementation that ensures a prearranged level of operational performance will be met during a contractual measurement period. High-availability features provide the ability of the user community to access the system, whether to submit new work, update or alter existing work, or collect the results of previous work.

5. **Namespace features**. These features represents special approaches and techniques that to make possible to represent data by means of hierarchy of files or in any other way.

6. **Synchronization features**. Synchronization refers to one of two distinct but related concepts: synchronization of processes, and synchronization of data. Synchronization features provide file system's internal techniques and architectural primitives are used to implement data synchronization.

7. **Network features**. The every Distributed File System (DFS) represents by itself a complex distributed system that can provide file system's services by means of some of specially designed network protocol. This class of features describes architectural solutions that to make file system services available by means of using different network-oriented technologies.

8. **Security features**. Computer security is the field that covers all the processes and mechanisms by which computer-based equipment, information and services are protected from unintended or unauthorized access, change or destruction. Security features characterize file system's approaches and techniques that to protect against data corruption or lost because of any malicious actions.

9. **Scalability features**. Scalability is the ability of a system, network, or process, to handle a growing amount of work in a capable manner or its ability to be enlarged to accommodate that growth.

## 3.1 Distributed File Systems

### 3.1.1 HDFS

HDFS [2] , [3] , [4]  is a file system designed for storing very large files with streaming data access patterns, running on clusters on commodity hardware. "Very large" in this context means files that are hundreds of megabytes, gigabytes, or terabytes in size.

HDFS is built around the idea that the most efficient data processing pattern is a write-once, read-many-times pattern. A dataset is typically generated or copied from source, and then various analyses are performed on that dataset over time. Each analysis will involve a large proportion, if not all, of the dataset, so the time to read the whole dataset is more important than the latency in reading the first record.

**Table 1 HDFS Features Classification**

| | HDFS |
|---|---|
| ARCHITECTURAL FEATURES | (1) Streaming Data Access Patterns; (2) Write-Once, Read-Many-Times Pattern; (3) Failure is a norm; (4) Single Writer – Many Readers; (5) Single NameNode in cluster; (6) DataNodes; (7) HDFS client; (8) NameNode is a multithreaded system; (9) Configurable Block Placement Policy Interface. |
| PERFORMANCE OPTIMIZATION FEATURES | (1) Namespace in RAM; (2) Scheduling a task to the data location; (3) Critical Files Replication Factor; (4) Batching of multiple transactions initiated by different clients; (5) MapReduce optimization; (6) DistCp tool. |
| RELIABILITY FEATURES | (1) Persistent Checkpoint; (2) File content replication; (3) Journal; (4) Redundant copies of the checkpoint and journal; (5) Data + [Checksum + Timestamp]; (6) NameNode initialization (checkpoint + journal); (7) Checkpoint and journal replication; (8) BackupNode; (9) BackupNode as read-only NameNode; (10) Snapshot; (11) Block scanner. |
| HIGH-AVAILABILITY FEATURES | (1) Critical Files Replication Factor; (2) CheckpointNode; (3) Periodical checkpointing; (4) Replication policy; (5) Over-replicated blocks policy; (6) Under-replicated blocks policy; (7) Background replication thread; (8) Balancer tool; (9) Corrupted blocks policy; (10) Decommissioned DataNode. |
| NAMESPACE FEATURES | (1) Hierarchy of files and directories; (2) Files and directories are represented on the NameNode by inodes; (3) The user references files and directories by paths in the namespace. |
| SYNCHRONIZATION FEATURES | (1) Write-Once, Read-Many-Times Pattern; (2) Single Writer – Many Readers; (3) File lease; (4) File change visibility. |
| NETWORK FEATURES | (1) File Content Replication; (2) Handshake; (3) Namespace ID; (4) Storage ID; (5) Block report; (6) Heartbeat; (7) DataNodes Write Pipeline; (8) Ordering by distance from the reader; (9) Rack's belonging; (10) Replicas distribution policy; (11) Decommissioning. |

**Figure 1 HDFS Features' Weight Comparison**

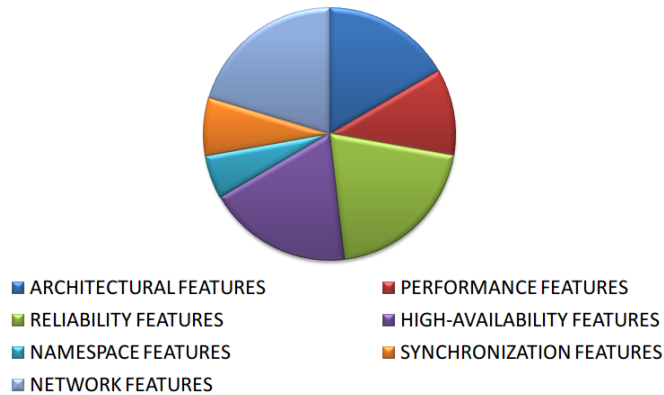

It is possible to see from the Figure 1 that the most significant classes of features in HDFS are: (1) Network features; (2) Reliability features; (3) High-availability features. The initial goal of this file system was reliable functioning on the basis of commodity hardware. Thereby, such feature as "Failure is a norm" was the base reason of such features' correlation.

### 3.1.2 GFS (Google File System)

GFS (Google File System) [5] , [6]  is designed as a distributed file system to be run on clusters up to thousands of machines. Running on commodity hardware, GFS is not only challenged by managing distribution, it also has to cope with the increased danger of hardware faults. Consequently, one of the assumptions made in the design of GFS is to consider disk faults, machine faults as well as network faults as being the norm rather than the exception.

GFS has been fully customized to suite Google's needs. GFS being targeted at a particular set of usage scenarios is optimized for usage of large files only with space efficiency being of minor importance. Moreover, GFS files are commonly modified by appending data, whereas modifications at arbitrary file offsets are rare. The majority of files can thus be considered as being append-only or even immutable (write once, read many).

GFS does not provide a POSIX interface. Instead, GFS implements a proprietary interface that applications can use.

**Table 2 GFS Features Classification**

| GFS (Google File System) | |
|---|---|
| ARCHITECTURAL FEATURES | (1) Write once, read many; (2) Typical workloads: Large Streaming Reads and Small Random Reads; (3) Producer-Consumer Queue; (4) Client side caching is not used; (5) Does not provide a POSIX interface; (6) A GFS cluster consists of a single master and multiple chunkservers and is accessed by multiple clients; (7) Files are divided into fixed-size chunks (64 MB); (8) Single Master in cluster; (9) Chunkservers store chunks on local disks as Linux files; (10) Lazy space allocation avoids wasting space due to internal fragmentation; (11) Persistent operation log - the namespaces and file-to-chunk mapping are kept persistent by logging mutations. |
| PERFORMANCE FEATURES | (1) Metadata in RAM; (2) Large files operation efficiency; (3) Large streaming reads; (4) Clients and chunkservers don't cache file data; (5) Clients do cache metadata; (6) Large chunk size reduces clients' need to interact with the master; (7) Persistent TCP connection to the chunkserver over an extended period of time; (8) Batch of several log records together before flushing on master side. |
| SYNCHRONIZATION FEATURES | (1) Concurrent records append operation; (2) Node, directory, file read-write locks; (3) Atomic file namespace mutation; (4) Stale replica processing; (5) Corrupted/Lost chunk processing; (6) GFS applications' relaxed consistency model; (7) Checkpointing; (8) Append-at-least-once semantic; (9) Chunk lease. |
| RELIABILITY FEATURES | (1) Failure is a norm; (2) Snapshot; (3) Chunk replication; (4) Operation log replication; (5) File system state recovering; (6) Checkpoints replication; (7) Checksumming on chunkservers; (8) Verifying inactive chunks; (9) Appending is preferable by random writes. |
| HIGH-AVAILABILITY FEATURES | (1) HeartBeat; (2) Periodical re-scan of chunkservers' entire state in the background; (3) Chunkservers report (startup + periodical reports); (4) Checkpointing by log size threshold; (5) Fast recovery; (6) Replication; (7) Master/Chunkserver fast restart; (8) "Shadow" master. |
| NAMESPACE FEATURES | (1) Per-directory data structure is absent; (2) Hard and symbolic links are absent; (3) Namespace is a lookup table. |
| NETWORK FEATURES | (1) Chunk replica placement policy; (2) The flow of data decouples from the flow of control; (3) Chunkservers pipeline; (4) Forwarding data to the "closest" machine; (5) Immediate data forwarding. |

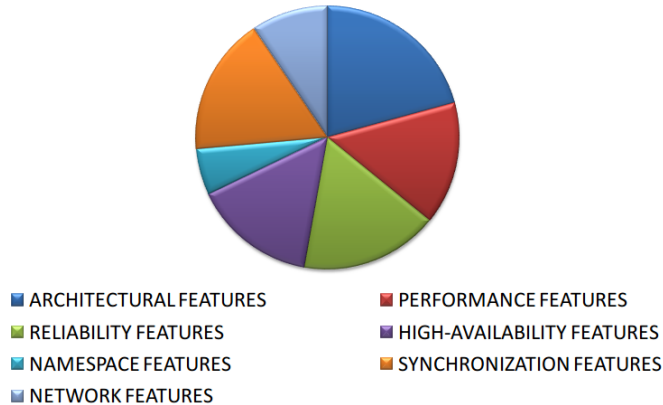

Figure 2 GFS Features' Weight Comparison

- ARCHITECTURAL FEATURES
- PERFORMANCE FEATURES
- RELIABILITY FEATURES
- HIGH-AVAILABILITY FEATURES
- NAMESPACE FEATURES
- SYNCHRONIZATION FEATURES
- NETWORK FEATURES

It is possible to see from the Figure 2 that the most significant classes of features in GFS are: (1) Synchronization features; (2) Reliability features; (3) High-availability features; (4) Performance optimization features. There two reasons for such features' correlation: (1) Necessity to be run on clusters up to thousands of machines with commodity hardware; (2) Trying to optimize file system for a particular set of usage scenarios.

### 3.1.3 InterMezzo

InterMezzo [7] , [8]  is a filtering file system layer, which sits in between the virtual file system and a specific file system such as ext3, tmpfs, ReiserFS, JFS, XFS. A partition or logical volume which is formatted as an InterMezzo file system is still a valid file system of the type that is being filtered, but InterMezzo adds a few files and directories for control purposes.

InterMezzo file system design is based on requirements:

1. The server file storage must reside in a native file system.
2. InterMezzo's client kernel level file system should exploit existing file systems, and have a persistent cache.
3. File system objects should have meta-data suitable for disconnected operation.
4. Scalability and recovery of the distributed state should leverage scalability and recovery of the local file systems.
5. The system should perform kernel level write back caching.
6. The system should use TCP and be designed to exploit existing advanced protocols such as rsync for synchronization and ssl/ssh for security.
7. Management of the client cache and server file systems should differ in policy, but use the same mechanisms.

**Table 3 InterMezzo Features Classification**

| **InterMezzo** | |
|---|---|
| ARCHITECTURAL FEATURES | (1) Filtering file system layer; (2) Fileset (subtree of the directory tree); (3) Kernel modification log (log of operations suitable for replay on other systems); (4) Received records file (records from remote replicas for reintegration); (5) Synchronization modification log (sync up an empty replica); (6) Disconnected operations; (7) Journaling file system; (8) Kernel Module (Presto); (9) User space cache manager (Lento). |
| PERFORMANCE FEATURES | (1) Caching of write operations; (2) Fetching of objects. |
| SYNCHRONIZATION FEATURES | (1) Synchronization after close; (2) Kernel based I/O-daemon; (3) Reintegration of KML from other fileset; (4) File system UUID; (5) Update notification; (6) Modification log; (7) Replication request; (8) Callback on object; (9) Object is identified by identifier and a version stamp; (10) Cached object validation; (11) Reintegrate request. |
| NAMESPACE FEATURES | (1) Fileset location database (FSLDB); (2) Root fileset. |
| SECURITY FEATURES | (1) ACLs; (2) Client can trust an authenticated server. |

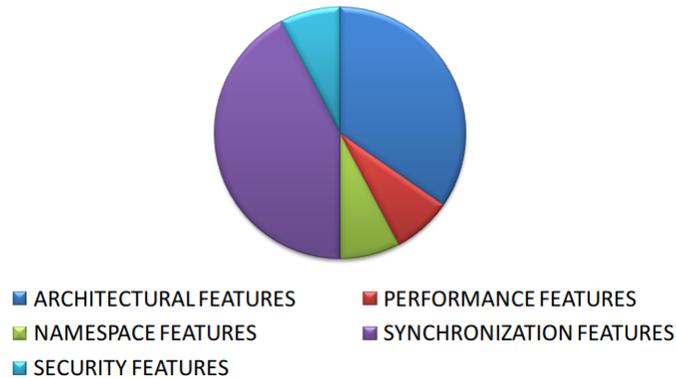

Figure 3 InterMezzo Features' Weight Comparison

### 3.1.4 Coda

Coda [9] , [10] , [11] , [12] , [13]  was designed to be a scalable, secure, and highly available distributed file system. An important goal was to achieve a high degree of naming and location transparency so that the system would appear to its users very similar to a pure local file system. By also taking high availability into account, the designers of Coda have also tried to reach a high degree of failure transparency.

Coda is designed for an environment consisting of a large collection of untrusted Unix clients and a much smaller number of trusted Unix file servers. The design is optimized for the access and sharing patterns typical of academic and research environments. It is specifically not intended for applications that exhibit highly concurrent, fine granularity data access.

Features of the Coda file system:
1. **Mobile Computing**: (1) Disconnected operation for mobile clients; (2) Reintegration of data from disconnected clients; (3) Bandwidth adaptation.

2. **Failure Resilience**: (1) Read/write replication servers; (2) Resolution of server/server conflicts; (3) Handles network failures which partition the servers; (4) Handles disconnection of client's client.
3. **Performance and scalability**: (1) Client-side persistent caching of files, directories and attributes for high performance; (2) Write-back caching.
4. **Security**: (1) Kerberos-like authentication; (2) Access control lists (ACLs).
5. **Well defined semantics of sharing**.

**Table 4 Coda Features Classification**

| Coda | |
|---|---|
| ARCHITECTURAL FEATURES | (1) Untrusted Unix client; (2) Trusted Unix file server; (3) Venus user-level process (provides access clients to files); (4) Vice file server; (5) Authentication server; (6) Update process; (7) Concurrent user-space threads. |
| PERFORMANCE FEATURES | (1) Cache manager (volume mappings); (2) Server replication by means of parallel access protocol; (3) Client-side persistent caching; (4) Write-back caching; (5) Cache coherence protocol; (6) Separate thread is used to handle all I/O operations; (7) Battery power-supplied write-ahead log. |
| SYNCHRONIZATION FEATURES | (1) Cache coherence protocol based on callbacks (callback break); (2) A pessimistic/optimistic approach towards disconnected operation; (3) A session is treated as a transaction; (4) Versioning scheme of file updates. |
| RELIABILITY FEATURES | (1) Single mechanism to cope with all disconnections; (2) First/Second class replica; (3) Server replication; (4) Cache entire files; (5) Recoverable Virtual Memory (RVM). |
| HIGH-AVAILABILITY FEATURES | (1) Server replication; (2) Client caches volume mappings; (3) Disconnected operation; (4) Reintegration of data from disconnected clients; (5) Resolution of server/server conflicts; (6) Handles network failures which partition the servers; (7) Pessimistic/Optimistic replication strategy; (8) File sharing by transferring of file's entire copy; (9) Read-One, Write-All (ROWA) protocol maintain consistency of a replicated volume; (10) Versioning scheme of file replication inconsistency detection and resolving; (11) Hoarding (filling the cache in advance); (12) Cache equilibrium (useful data are indeed cached). |
| NAMESPACE FEATURES | (1) Volume (partial subtree in the shared name space keeps a collection of files associated with a user); (2) Possibility to operate for extended periods in isolation; (3) Globally shared name space; (4) UNIX-like naming system; (5) Mounting point (a leaf node of a volume that refers to the root node of another volume); (6) Automatic mount a volumes during name lookup; (7) Replicated Volume Identifier (RVID); (8) File identifier (RVID + file handle). |
| NETWORK FEATURES | (1) Bandwidth adaptation; (2) User-level RPC system (communication channel between client and server); (3) RPC2 (reliable RPCs on top of the [unreliable] UDP protocol); (4) Side effects (client and server can communicate using an application-specific protocol); (5) Multicasting (send an invalidation message to all clients in parallel); (6) MultiRPC (Parallel RPCs). |
| SECURITY FEATURES | (1) Access control lists (ACLs); (2) Secure RPC channel between a client and a server; (3) Secure login session (Kerberos-like authentication); (4) User/Group rights. |
| SCALABILITY FEATURES | (1) Callback-based cache coherence; (2) Whole-file caching; (3) Placing of functionality on clients rather than servers; (4) Avoidance of system-wide rapid change; (5) Client-side caching. |

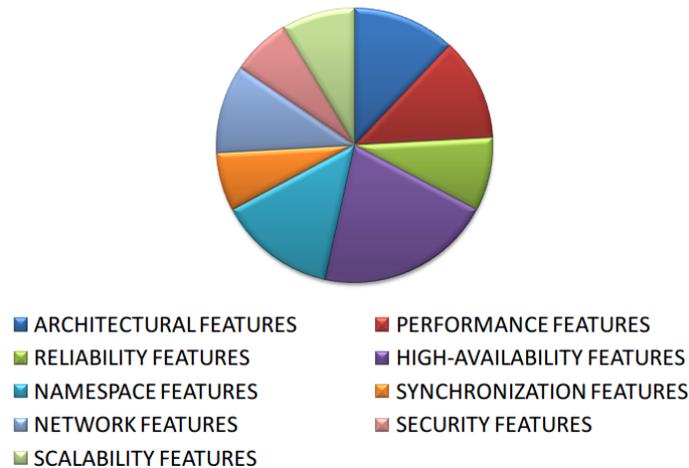

Figure 4 Coda Features' Weight Comparison

It is possible to see from the Figure 4 that the most significant classes of features in Coda are: (1) High-availability features; (2) Namespace features; (3) Performance optimization features; (4) Network features. The core feature of the Coda is disconnected operations and, therefore, it is the reason of such features' correlation.

### 3.1.5 Ceph

Ceph [14] is a distributed file system that architecture is based on the assumption that systems at the petabyte scale are inherently dynamic: large systems are inevitably built incrementally, node failures are the norm rather than the exception, and the quality and character of workloads are constantly shifting over time.

Ceph decouples data and metadata operations by eliminating file allocation tables and replacing them with generating functions. This allows Ceph to leverage the intelligence present in OSDs to distribute the complexity surrounding data access, update serialization, replication and reliability, failure detection, and recovery. Ceph utilizes a highly adaptive distributed metadata cluster architecture that improves the scalability of metadata access, and with it, the scalability of the entire system.

**Table 5 Ceph Features Classification**

| Ceph | |
|---|---|
| ARCHITECTURAL FEATURES | (1) Systems at the petabyte scale are inherently dynamic; (2) Decouple data and metadata operations; (3) Adaptive distributed metadata cluster architecture; (4) Clients; (5) Near-POSIX file system interface; (6) Cluster of OSDs (Object Storage Devices); (7) Metadata servers cluster; (8) Dynamic distributed metadata management; (9) Reliable Autonomic Distributed Object Store (RADOS); (10) Extent and B-tree based Object File System (EBOFS). |
| PERFORMANCE FEATURES | (1) Controlled Replication Under Scalable Hashing (CRUSH); (2) Dynamic hierarchical partition; (3) Optimization for the most common metadata access scenarios; (4) No file allocation metadata is necessary; (5) In-memory cache; (6) Lazily flushed journals strategy; (7) Inodes embedded in directory; (8) Partitioning the directory hierarchy across multiple nodes; (9) Knowledge of metadata popularity; (10) Hashing content of large or heavy load directories by file name across the cluster; (11) Specially optimized low-level disk scheduler; (12) B-tree service of EBOFS. |
| SYNCHRONIZATION FEATURES | (1) Object locks; (2) O_LAZY flag (allows applications to explicitly relax the usual coherency requirements for a shared-write file); (3) No metadata locks or leases are issued to clients; (4) Capabilities (specifying which operations are permitted); (5) Shared long-term storage and carefully constructed namespace locks. |
| RELIABILITY FEATURES | (1) Commit metadata updates to disk; (2) Lazily flushed journals of MDS; (3) Quick rescan of MDS journal by any node in the case of MDS failure; (4) OSDs self-report; (5) Active monitoring of OSDs peers in PG; (6) OSD liveness (OSD reachable + assigning data by CRUSH); (7) Object version number + PG's log of recent changes; (8) Fast Recovery Mechanism (FaRM). |
| HIGH-AVAILABILITY FEATURES | (1) Cluster of OSDs; (2) Uniform striping and distribution strategy; (3) Placement Groups (PG) + Controlled Replication Under Scalable Hashing (CRUSH); (4) Failure is the norm; (5) Primary-copy replication; (6) OSD monitor. |
| NAMESPACE FEATURES | (1) CRUSH; (2) Dynamic Subtree Partitioning; (3) MDS cluster; (4) Object names simply combine the file inode number and the stripe number; (5) Directory's content distribution strategy; (6) Ranges of inode numbers; (7) Auxiliary anchor table (rare inode with multiple hard links); (8) Single authoritative MDS; (9) Popularity of metadata; (10) Three groups of inode contents with different consistency semantics (security, file, and immutable). |
| SECURITY FEATURES | (1) Capabilities (specifying which operations are permitted). |
| SCALABILITY FEATURES | (1) Object-based storages; (2) Object names are constructed using the inode number, and distributed to OSDs using CRUSH; (3) MDS response content; (4) Future metadata operations are directed at the authority (for updates) or a random replica (for reads) based on the deepest known prefix of a given path. |

**Figure 5 Ceph Features' Weight Comparison**

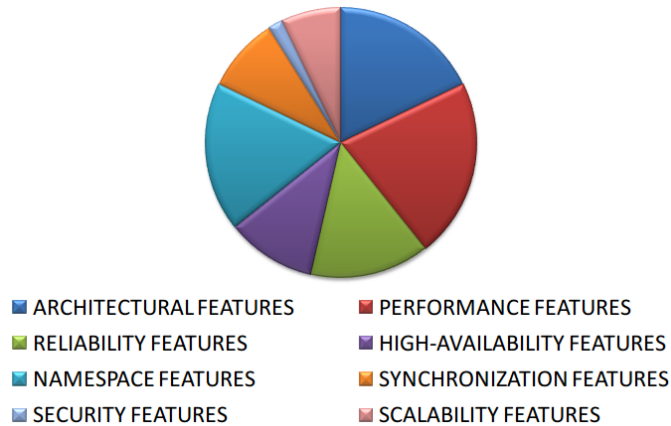

- ARCHITECTURAL FEATURES
- PERFORMANCE FEATURES
- RELIABILITY FEATURES
- HIGH-AVAILABILITY FEATURES
- NAMESPACE FEATURES
- SYNCHRONIZATION FEATURES
- SECURITY FEATURES
- SCALABILITY FEATURES

It is possible to see from the Figure 5 that the most significant classes of features in Ceph are: (1)

Performance optimization features; (2) Namespace features; (3) Reliability features; (4) High-availability features; (5) Synchronization features. The main goal of the file system is to improve the scalability of metadata access, and with it, the scalability of the entire system. Thereby, it is resulted in such extent of features' correlation.

### 3.1.6  DDFS

Disco Distributed Filesystem (DDFS) [15] provides a distributed storage layer for Disco (Disco is a distributed computing framework based on the MapReduce paradigm). DDFS is designed specifically to support use cases that are typical for Disco and MapReduce in general: storage and processing of massive amounts of immutable data. DDFS is complementary to traditional relational databases or distributed key-value stores, which often have difficulties in scaling to tera- or petabytes of bulk data. It is not a general-purpose POSIX-compatible filesystem.

DDFS is a low-level component in the Disco stack, taking care of data distribution, replication, persistence, addressing and access. It does not provide a sophisticated query facility in itself but it is tightly integrated with Disco jobs. Disco can store job results to DDFS, providing persistence for and easy access to processed data.

DDFS is a tag-based filesystem: instead of having to organize data to directory hierarchies, you can tag sets of objects with arbitrary names and retrieve them later based on the given tags. Tags can contain links to other tags, and data can be referred to by multiple tags; tags hence form a network or a directed graph of metadata. DDFS is schema-free, so you can use it to store arbitrary, non-normalized data.

DDFS is designed to operate on commodity hardware. Fault-tolerance and high availability are ensured by K-way replication of both data and metadata, so the system tolerates K-1 simultaneous hardware failures without interruptions. DDFS stores data and metadata on normal local filesystems, such as ext3 or xfs, so even under a catastrophic failure data is recoverable using standard tools.

**Table 6 DDFS Features Classification**

| DDFS | |
|---|---|
| ARCHITECTURAL FEATURES | (1) Not POSIX-compatible; (2) Store and process of massive amounts of immutable data; (3) Low-level component in the Disco stack; (4) Schema-free (store arbitrary, non-normalized data); (5) Not suitable for storing very small (fewer than 4K) data; (6) Blob; (7) Tag; (8) Single master node; (9) Storage node; (10) Client; (11) Master cache; (12) Tag server; (13) Background garbage collection and re-replication process. |
| PERFORMANCE FEATURES | (1) Load balancing; (2) Cache of all tags stored on the node. |
| SYNCHRONIZATION FEATURES | (1) Token (atomic creation of access-controlled tags); (2) All tag-related operations are handled by the master to ensure their atomicity and consistency; (3) Modified 3-phase commit protocol; (4) Message limitation by timeout; (5) The +deleted list for deleted tags. |
| RELIABILITY FEATURES | (1) Journaled File System. |
| HIGH-AVAILABILITY FEATURES | (1) K-way replication of both data and metadata; (2) Blobs replication; (3) Tags replication; (4) Garbage collection process; (5) Nodes availability monitor; (6) Node space partitioning using a consistent hashing mechanism. |
| NAMESPACE FEATURES | (1) Tag-based filesystem; (2) Directed graph of metadata; (3) Store arbitrary attributes with the tags; (4) Metadata is handled centrally; (5) Orphaned blobs; (6) Alternative views to the same data; (7) User-defined attributes. |
| NETWORK FEATURES | (1) HTTP or direct access to blobs. |
| SECURITY FEATURES | (1) Token-based authorization mechanism. |
| SCALABILITY FEATURES | (1) Storage nodes can be added to the cluster on the fly. |

**Figure 6 DDFS Features' Weight Comparison**

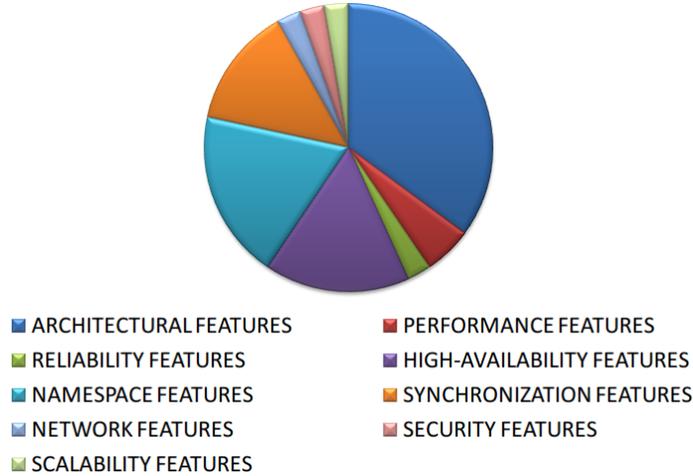

It is possible to see from the Figure 6 that the most significant classes of features in DDFS are: (1) Namespace features; (2) High-availability features; (3) Synchronization features. It is possible to interpret such features' correlation by means of aim to realize a tag-based file system in the environment of commodity hardware.

### 3.1.7 zFS

zFS [16] , [17] is a research project aimed at building a decentralized file system that distributes all aspects of file and storage management over a set of cooperating machines interconnected by a high-speed network. zFS is designed to be a file system that will (1) Scale from a few networked computers to several thousand machines, supporting tens of thousands of clients and (2) Be built from commodity, off-the-shelf components such as PCs, Object Store Devices (OSDs) and a high-speed network, and run on existing operating systems such as Linux.

The design and implementation of zFS is aimed at achieving a scalable file system. The objectives of zFS are:

1. Creating a file system that operates equally well on few or thousands of machines.
2. Using off-the-shelf components with OSDs.
3. Making use of the memory of all participating machines as a global cache to increase performance.
4. Achieving almost linear scalability: the addition of machines will lead to an almost linear increase in performance.

zFS achieves scalability by separating storage management from file management and by dynamically distributing file management. Storage management in zFS is encapsulated in the Object Store Devices (OSDs), while file management is done by other zFS components. Having OSDs handle storage management implies that functions usually handled by file systems are done in the OSD itself, and are transparent to other components of zFS. These include: data striping, mirroring, and continuous copy/PPRC. The Object Store does not distinguish between files and directories. It is the responsibility of the file system management (the other components of zFS) to handle them correctly.

**Table 7 zFS Features Classification**

| zFS | |
|---|---|
| ARCHITECTURAL FEATURES | (1) Object Store Devices (OSDs); (2) zFS front-end (provides access to zFS files and directories). |
| PERFORMANCE FEATURES | (1) Coherent cache (cooperative cache); (2) Data blocks are sent to the OSD asynchronously; (3) Every opened file is managed by single file manager; (4) Directory operations become distributed transactions; (5) zFS pre-fetching mechanism (correction of read-ahead). |
| RELIABILITY FEATURES | (1) Distributed transactions; (2) Transaction server (TSVR). |
| HIGH-AVAILABILITY FEATURES | (1) Data striping, mirroring, and continuous copy/PPRC; (2) Singlet/Replicated data blocks. |
| NAMESPACE FEATURES | (1) Storage objects form flat ID space; (2) File object; (3) Directory object; (4) Higher level management and copy-services provided by the OSD; (5) File manager (FMGR); (6) zFS uses the object-stores to lay out both files and directories. |
| SYNCHRONIZATION FEATURES | (1) Lease manager (grants the file managers exclusive file leases on whole file); (2) File manager (each lease request for any part of the file F is handled by associated file manager); (3) Lease; (4) OSD stores in memory the network address of the current holder of the major-lease. |
| SECURITY FEATURES | (1) The object store provides security enforcement for access to the storage-objects it contains but it does not provide security management; (2) Securing all operations with a credential; (3) Provide increased protection/security at the level of objects rather than whole LUs; (4) Object store security model. |
| SCALABILITY FEATURES | (1) Separating storage management from file management; (2) Object store control unit may export multiple object stores; (3) Distributing space allocation among storage controllers; (4) The |

| | cooperative cache is a key component in achieving high scalability. |
|---|---|

**Figure 7 zFS Features' Weight Comparison**

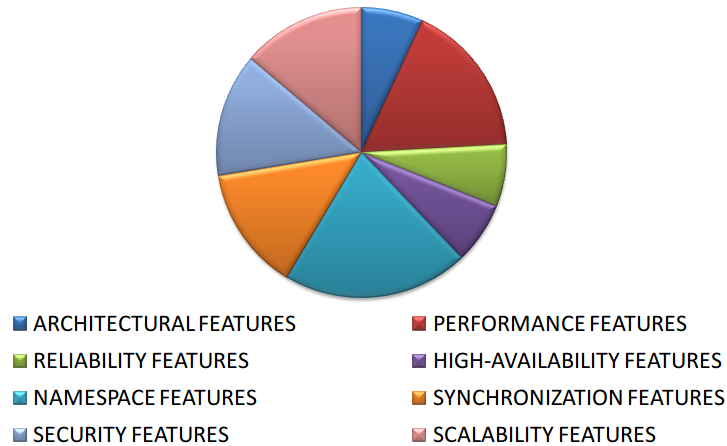

It is possible to see from the Figure 7 that the most significant classes of features in zFS are: (1) Namespace features; (2) Performance optimization features; (3) Security features; (4) Scalability features; (5) Synchronization features. The goal to be a decentralized and scalable file system is resulted in significant efforts in the direction of namespace, security and scalability approaches. Moreover, using a high-speed network as a basis gives opportunities for such performance optimization technique as coherent cache.

### 3.1.8 Zebra

Zebra [18] is designed to provide a file transfer rate that scales with the number of servers. Zebra increases throughput by striping file data across multiple servers and it increases availability and reliability by using parity to mask single server failures. Instead of striping individual files, Zebra forms all the new data from each client into a stream, which is then striped across the servers. This allows the data from many small writes to be batched together and stored on a server in a single transfer, reducing the per-file overhead and improving server efficiency.

Zebra makes several assumptions concerning its computing environment and the types of failures that it will withstand. Zebra is designed to support UNIX workloads as found in office/engineering environments. These workloads are characterized by short file lifetimes, sequential file accesses, infrequent write-sharing of a file by different clients, and many small files. Zebra is therefore designed to handle sequential file accesses well, perhaps at the expense of random file accesses.

Zebra is also targeted at high-speed local-area networks. Zebra is not designed specifically to reduce network traffic. It is assumed that in a data transfer between a client and server the point-to-point bandwidth of the network is not a bottleneck. Zebra is also not designed to handle network partitions.

Zebra also assumes that clients and servers will have large main-memory caches to store file data. These caches serve two purposes: to allow frequently used data to be buffered and accessed in memory, without requiring an access to the server or the disk; and to buffer newly written file data prior to writing it to the server or the disk.

Zebra is designed to provide file service despite the loss of any single machine in the system. Multiple server failures are not handled.

**Table 8 Zebra Features Classification**

| Zebra | |
|---|---|
| ARCHITECTURAL FEATURES | (1) Support UNIX workloads; (2) Scalable filesystem; (3) Targeted at high-speed local-area networks; (4) Not designed to handle network partitions; (5) Stripe cleaner; (6) Log-based striping; (7) Client; (8) Storage server; (9) File manager; (10) Data fragment; (11) Parity fragment. |
| PERFORMANCE FEATURES | (1) Striping file data across multiple servers by means of stream; (2) Large main-memory caches; (3) Log-based striping; (4) Batching many small writes by application programs into full stripe write; (5) Virtual stripe; (6) Append-only writes + virtual stripes; (7) Use of log addresses to access data on the servers; (8) Large files read-ahead; (9) Prefetch small files by reading entire stripe at a time; (10) Zebra clients use write-back caches; (11) Client should transfer fragments to all of the storage servers concurrently; (12) Batching of small data pieces for the case of frequent fsync; (13) Caching block pointers on client; (14) Caching naming information on clients; (15) Multiple file managers. |
| RELIABILITY FEATURES | (1) Parity is used to mask server failures; (2) Stripe log; (3) Delta (describe changes in the state of the file system); (4) Write-ahead logging (A write-ahead log is used to record a set of actions that are about to be performed, before actually performing them); (5) All storage server operations are synchronous and atomic; (6) Fragment checksum; (7) Sequence numbers associated with parity fragments; (8) Recovering storage server; (9) Stripe fragment reconstruction; (10) Log can be used as reliable communication channel. |
| HIGH-AVAILABILITY FEATURES | (1) Parity is used to mask server failures. |
| NAMESPACE FEATURES | (1) Single file manager; (2) Inode with log addresses; (3) Central synchronization point for metadata modifications. |
| SYNCHRONIZATION FEATURES | (1) If several clients simultaneously modify the blocks contained in a single stripe, they can write their modified blocks to different stripes and avoid the synchronization; (2) Virtual stripes; (3) Block maps (keep track of where each file block is located); (4) Cache consistency protocol; (5) File manager as a centralized service; (6) Log as reliable communication channel; (7) Delta (describe changes in the state of the file system); (8) Cache consistency mechanism. |

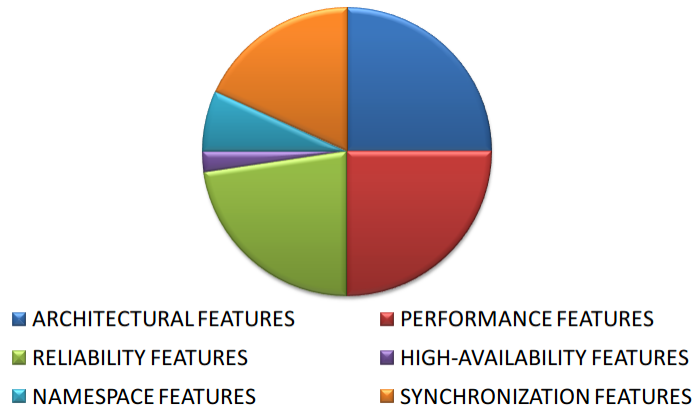

Figure 8 Zebra Features' Weight Comparison

It is possible to see from the Figure 8 that the most significant classes of features in Zebra are: (1) Performance optimization features; (2) Reliability features; (3) Synchronization features. The main goal of design was to provide a file transfer rate that scales with the number of servers. Thereby, performance optimization features were the main direction of design efforts. The striping is the cornerstone of the design at whole. This approach gives opportunities for evolution of reliability and synchronization approaches in special internal techniques (stripe log, using parity for masking server failure, virtual stripes).

### 3.1.9 PlasmaFS

PlasmaFS [19], [20] is a distributed filesystem for large files, implemented in user space. PlasmaFS is deployed on an arbitrary number of NameNodes and DataNodes. All data and metadata is replicated. ACID transactions (the ACID properties are atomicity, consistency, isolation, and durability) provide data safety and clear query semantics. PlasmaFS focuses on large files and blocksizes in the range 64K to 1M. It is error-resilient and extensible.

PlasmaFS uses a data store with full transactional support (PostgreSQL). PlasmaFS provides a transactional view to users. This works very much like the transactions in SQL. The performance advantage is here that several write operations can be carried out with only one commit. PlasmaFS takes it that far that unlimited numbers of metadata operations can be put into a transaction, such as creating and deleting files, allocating blocks for the files, and retrieving block lists.

A version number is maintained per file that is increased whenever data or metadata are modified. This allows it to keep external caches up to date with only low overhead: A quick check whether the version number has changed is sufficient to decide whether the cache needs to be refreshed.

**Table 9 PlasmaFS Features Classification**

| PlasmaFS | |
|---|---|
| ARCHITECTURAL FEATURES | (1) User-space implementation; (2) NameNodes; (3) DataNodes; (4) Metadata is stored in PostgreSQL databases; (5) Master (coordinator) NameNode; (6) Inodecache NameNode server; (7) EOF position in the inodeinfo struct. |
| PERFORMANCE FEATURES | (1) Inodecache NameNode server; (2) Transactional view (unlimited numbers of metadata operations can be put into a transaction); (3) PlasmaFS addresses blocks linearly (extent-based scheme); (4) A version number is maintained per file; (5) High parallelism of DB transactions; (6) Write serialization; (7) Datanode program can handle multiple requests simultaneously. |
| RELIABILITY FEATURES | (1) ACID (atomicity, consistency, isolation, and durability) transaction; (2) Data store with full transactional support (PostgreSQL); (3) Two-phase commit; (4) All metadata and data accesses are done in a transactional way; (5) Several simultaneous transactions on the same TCP connection; (6) Replacement blocks for the file region allocation. |
| HIGH-AVAILABILITY FEATURES | (1) Replication; (2) Replication is ACID-compliant. |
| NAMESPACE FEATURES | (1) PlasmaFS protocol returns the inode ID's to the user, and the user can also access files by this ID; (2) Inode objects (metadata + content data); (3) Sequence number in inodeinfo (a version number of the contents); (4) PlasmaFS stores directories in the namenode database; (5) Special PostgreSQL table storing the directory tree; (6) Directory inode is used for storing access rights; (7) Inode can has several file names (likewise hardlinks). |
| SYNCHRONIZATION FEATURES | (1) Ticket system; (2) Replacement blocks for the file region allocation; (3) Cryptographic scheme. |
| NETWORK FEATURES | (1) PlasmaFS uses SunRPC for all TCP connections. |

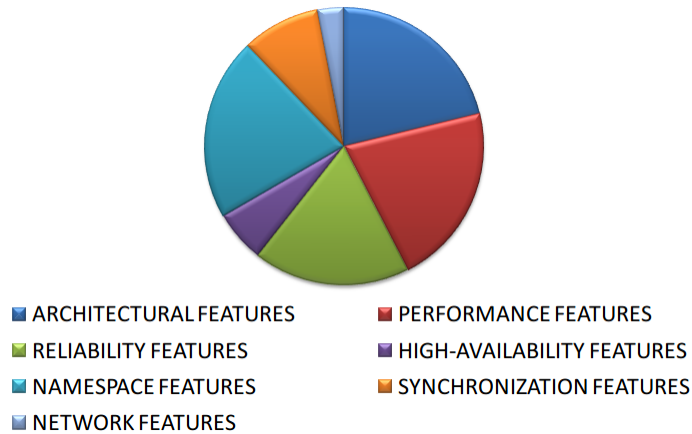

**Figure 9 PlasmaFS Features' Weight Comparison**

It is possible to see from the Figure 9 that the most significant classes of features in PlasmaFS are: (1) Namespace features; (2) Performance optimization features; (3) Reliability features. The core design's peculiarity is an ACID transactions and using a data store with full transactional support (PostgreSQL). This approach defines directions of features classes' evolution. As a result, it was suggested special namespace and reliability features' approaches. The data store is used as a cornerstone approach of performance optimization features.

## 3.2 Shared Storage Area Network (SAN) File Systems

### 3.2.1 XSAN

Xsan [21] , [22] , [23] , [24]  is a high-performance storage area network (SAN) file system for Mac OS X and Mac OS X Server. It enables users to share centralized disk storage with multiple computers over Fibre Channel.

Xsan is a 64-bit cluster file system specifically designed for small and large computing environments that demand the highest level of data availability. This specialized technology enables multiple Mac desktop and Xserve systems to share RAID storage volumes over a high-speed Fibre Channel network. Each client can read and write directly to the centralized file system, accelerating user productivity while improving workgroup collaboration.

**Table 10 XSAN Features Classification**

| XSAN | |
|---|---|
| ARCHITECTURAL FEATURES | (1) SAN, (2) Fibre Channel, (3) RAID, (4) "Metadata controller" |
| PERFORMANCE FEATURES | (1) SAN bandwidth reservation, (2) Multiple RAID devices pool, (3) Pretuned volume workload settings, (4) Storage affinities, (5) Real-time I/O mode |
| RELIABILITY FEATURES | (1) Journaled file system, (2) MultiSAN |
| HIGH-AVAILABILITY FEATURES | (1) Metadata controller failover, (2) Fibre Channel multipathing, (3) Standby controller, (4) Load-balancing |
| SYNCHRONIZATION FEATURES | (1) File-level locking |
| NETWORK FEATURES | (1) Fibre Channel switch, (2) AFP, (3) SMB/CIFS, (4) NFS |
| SECURITY FEATURES | (1) LDAP, (2) Flexible file permissions, (3) Storage volumes visibility |
| SCALABILITY FEATURES | (1) Flexible volume management |

**Figure 10 XSAN Features' Weight Comparison**

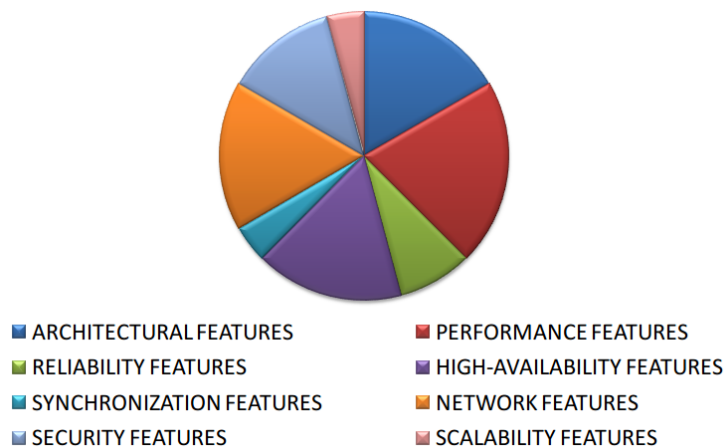

It is possible to see from the Figure 10 that the most significant classes of features in XSAN are: (1) Performance optimization features; (2) Network features; (3) High-availability features; (4) Security features.

### 3.2.2 CXFS

CXFS [25] , [26] , [27] , [28]  is a shared XFS filesystem that allows groups of computers to coherently share large amounts of data while maintaining high performance. It runs on storage area network (SAN) disks, such as Fibre Channel, in a cluster environment.

CXFS provides a single-system view of the filesystems; each host in the SAN has equally direct access to the shared disks and common path names to the files.

CXFS is a clustered XFS filesystem that allows for logical file sharing. CXFS runs on top of a storage area network (SAN), where each computer system in the cluster has direct high-speed data channels to a shared set of disks.

**Table 11 CXFS Features Classification**

| | CXFS |
|---|---|
| ARCHITECTURAL FEATURES | (1) Storage area network (SAN); (2) Cluster; (3) Pool; (4) Cluster membership. |
| PERFORMANCE FEATURES | (1) Storage area network (SAN); (2) Memory-mapped files; (3) Peer-to-disk model for the data access; (4) Advanced buffering techniques of the XFS filesystem; (5) Asynchronous buffering technique; (6) Multiple Host Bus Adapters; (7) Fast metadata lookups; (8) Ability to allocate large extents; (9) RPC design that enables high speed processing of metadata transactions; (10) Guaranteed Rate I/O; (11) Only the metadata passes through the server; data reads and writes are direct to disk; (12) Dedicated TCP/IP network for metadata. |
| RELIABILITY FEATURES | (1) Fault isolation and recovery; (2) Recovery of a log-based filesystem; (3) RAID arrays; (4) Journaling of the XFS log based filesystem. |
| HIGH-AVAILABILITY FEATURES | (1) Backup metadata servers; (2) Remote hardware reset in the case of software or hardware failure; (3) Cluster membership quorum; (4) Heartbeating; (5) Redundant Fibre Channel fabric; (6) Dedicated TCP/IP network for metadata and other CXFS traffic; (7) Journaled filesystem. |
| NAMESPACE FEATURES | (1) Single-system view of the filesystems (normal POSIX path name); (2) XFS directory structure is based on B-trees; (3) Metadata includes information about files as well as information about the filesystem itself; (4) Metadata transactions are routed over a TCP/IP network to the metadata server; (5) Metadata server; (6) Client/server model. |
| SYNCHRONIZATION FEATURES | (1) POSIX-compliant file locking; (2) Token mechanism; (3) Metadata server acts as a central clearinghouse for metadata logging, file locking, buffer coherency, and other necessary coordination functions. |
| NETWORK FEATURES | (1) Dedicated LAN for metadata traffic. |

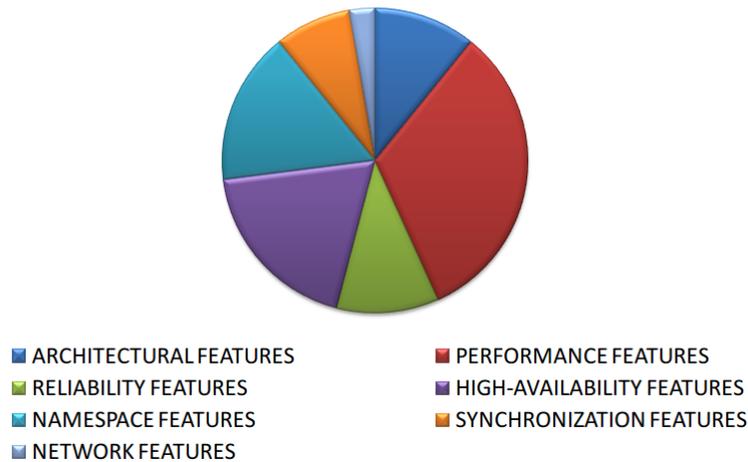

**Figure 11 CXFS Features' Weight Comparison**

It is possible to see from the Figure 11 that the most significant classes of features in CXFS are: (1) Performance optimization features; (2) High-availability features; (3) Namespace features. The background of the whole CXFS architecture is storage area network (SAN) disks and XFS's internal techniques.

## *3.3 Shared Cluster File Systems*

### 3.3.1 ACFS

Oracle Automatic Storage Management Cluster File System (Oracle ACFS) [29] is a multi-platform, scalable file system, and storage management technology that extends Oracle Automatic Storage Management (Oracle ASM) functionality to support customer files maintained outside of Oracle Database.

Oracle ACFS is tightly coupled with Oracle Clusterware technology, participating directly in Clusterware cluster membership state transitions and in Oracle Clusterware resource-based high availability (HA) management.

Oracle ACFS can be accessed and managed using native operating system file system tools and standard application programming interfaces (APIs). In addition to sharing file data, Oracle ACFS provides additional storage management services including support for the Oracle Restart mount registry and the Oracle Grid Infrastructure clusterwide mount registry, dynamic on-line file system resizing, and multiple space-efficient snapshots for each file system.

**Table 12 ACFS Features Classification**

| ACFS | |
|---|---|
| ARCHITECTURAL FEATURES | (1) Volume manager; (2) Native operating system file system application programming interfaces (APIs); (3) Disk group. |
| PERFORMANCE FEATURES | (1) Direct access to Oracle ASM disk group storage; (2) I/O parallelism; (3) Optimized fast directory lookup for large directories; (4) Collection of disks are managed as a unit; (5) Load balancing among all of the disks in a disk group; (6) Large size of Allocation Unit; (7) The contents of files are stored in a disk group as a set, or collection, of extents that are stored on individual disks within disk groups; (8) The initial extent size equals the disk group allocation unit size and it increases by a factor of 4 or 16 at predefined thresholds.; (9) Coarse-grained and fine-grained striping; (10) Rebalancing a disk group; (11) Multipathing; (12) Variable extent-based storage allocation; (13) Preallocation of large user files to improve performance when writing data. |
| RELIABILITY FEATURES | (1) Mirroring protection mechanism; (2) Redundancy level for each file; (3) Failure groups; (4) Metadata checksums and journaling; (5) Snapshot; (6) Replication; (7) Replication log. |
| HIGH-AVAILABILITY FEATURES | (1) Clusterwide mount registry; (2) Dynamic on-line file system resizing; (3) Multiple snapshots for each file system; (4) Cluster Synchronization Service; (5) Automatic redistribution the file contents in the case of adding or removing disks from a disk group; (6) Multipathing. |
| NAMESPACE FEATURES | (1) Clusterwide naming of all customer application files; (2) Clusterwide user and metadata cache coherency mechanism; (3) Hierarchical tree-structured namespace; (4) ACFS file system may be mounted into the native operating system file system namespace; (5) Tagging assigns a common naming attribute to a group of files. |
| NETWORK FEATURES | (1) NFS; (2) CIFS. |
| SECURITY FEATURES | (1) Realm-based security (group of files or directories that are secured for access by a user or a group of users); (2) Security rules; (3) Command rules; (4) ACLs; (5) Encryption. |

**Figure 12 ACFS Features' Weight Comparison**

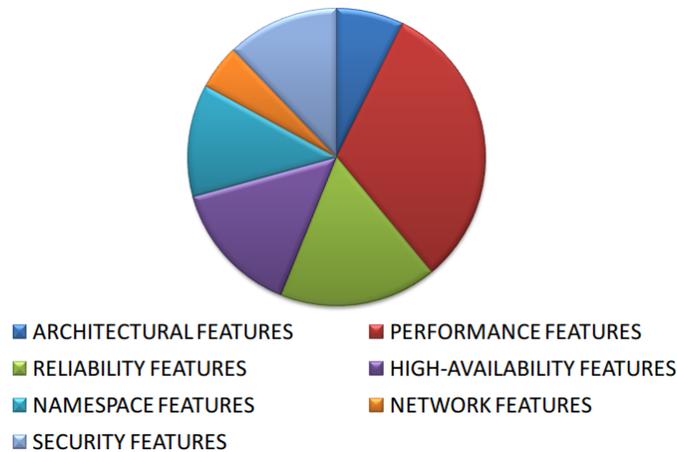

It is possible to see from the Figure 12 that the most significant classes of features in ACFS are: (1) Performance optimization features; (2) Reliability features; (3) High-availability features; (4) Namespace features; (5) Security features.

## *3.4 Clustered Distributed File Systems*

### 3.4.1 ExaFS

From the client's point of view, the ExaStore NAS [30] solution is presented as a single server with a single file system, IP address, and name, regardless of the number of nodes and storage subsystems.

ExaStore features:

1. Single global file name space that greatly simplifies information sharing.
2. Fully distributed file system that enables flexibility and agility.
3. Massive scalability supports growth in both capacity and bandwidth independently.
4. Fully clustered solutions that allow high availability, reliability, and reduces downtime.
5. Multi protocol compatibility allows administrators to consolidate many individual file servers into a single entity, which is accessible via all major network-file protocols, such as NFS, CIFS and AFP.

**Table 13 ExaFS Features Classification**

| ExaFS | |
|---|---|
| ARCHITECTURAL FEATURES | (1) Node; (2) RAID; (3) LUN; (4) Fibre Channel; (5) Interconnect Network; (6) Interconnect Switches. |
| PERFORMANCE FEATURES | (1) NVRAM; (2) Cache + Data-placement schemes; (3) Cache + Write to file; (4) Cache + Read file; (5) Idle Resource Utilization; (6) Fibre Channel; (7) Interconnect Network; (8) Interconnect Switches: Dual link + Load balancing; (9) Dedicated FSD's metadata file; (10) Traffic Load Balancing. |
| RELIABILITY FEATURES | (1) RAID; (2) Interconnect Switches; (3) Cache mirroring; (4) Dual access to storage array; (5) LUNs control failover; (6) FSD failover; (7) Journaling approach in degraded mode; (8) UPS notifications. |
| HIGH-AVAILABILITY FEATURES | (1) Node Pairs; (2) Interconnect Network; (3) Data distribution between LUNs; (4) LUNs control failover; (5) FSDs (file system daemons); (6) Dedicated FSD's metadata file; (7) Components redundancy; (8) Automatic Recovery; (9) Automatic reboot and shutdown of nodes. |
| NETWORK FEATURES | (1) NFS, CIFS, AFP; (2) Public network; (3) Virtual IP; (4) Traffic Load Balancing; (5) Dedicated VIPs; (6) Private Management Network; (7) System boot through the management network; (8) Multiple Network Paths. |
| SCALABILITY FEATURES | (1) Array Management (LUN expansions and additions); (2) Load Balancing Mechanism. |

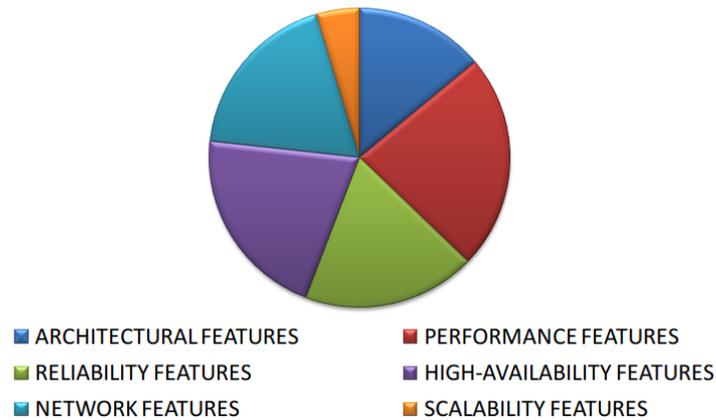

Figure 13 ExaFS Features' Weight Comparison

It is possible to see from the Figure 13 that the most significant classes of features in ExaFS are: (1) Performance optimization features; (2) High-availability features; (3) Reliability features; (4) Network features.

## *3.5 Wide Area File Systems*

### 3.5.1 Gfarm

Gfarm [31] , [32]  is a part of the Grid Datafarm architecture for Petascale data-intensive computing facilitating distributed resources in wide area. Main features of the architecture are to provide (1) a Grid file system that integrates local disks of compute nodes in Computational Grid, and (2) parallel and distributed computing associating Computational Grid and Data Grid.

A Grid file system is a global virtual file system that federates numbers of file systems (or file servers) in a Grid. Integration is achieved by a filesystem metadata server that manages a virtual human-readable namespace. As such, a Grid file system is a shared network file system scaled to Grid level, allowing easy and transparent sharing of file data without any modifications to existing applications.

Grid Datafarm architecture, moreover, supports high-performance distributed and parallel computing for processing a group of files by a single program, which is a most time-consuming, but also a most typical, task in data-intensive computing such as high energy physics, astronomy, space exploration, and human genome analysis. In order to facilitate this, in Grid Datafarm, an arbitrary group of files possibly dispersed across administrative domains can be managed by a single Gfarm file.

Gfarm v2 aims to provide a POSIX-compliant global virtual file system facilitating features of Grid Datafarm architecture for Petascale data-intensive computing. It can be used as a general-purpose network file system for Grid or virtual organization, allowing existing applications to share files securely and dependably, and to access files efficiently across administrative domains.

**Table 14 Gfarm Features Classification**

| | Gfarm |
|---|---|
| ARCHITECTURAL FEATURES | (1) Computational Grid; (2) Data Grid; (3) Grid file system; (4) POSIX-compliant global virtual file system; (5) Read-write file open mode and advisory file locking; (6) Parallel data processing capability; (7) Gfarm I/O library; (8) Gfsd (I/O daemon); (9) Metadata server (LDAP server). |
| PERFORMANCE FEATURES | (1) File-affinity process scheduling; (2) File view. |
| NAMESPACE FEATURES | (1) File view; (2) System call hooking library to access Gfarm file system. |
| SYNCHRONIZATION FEATURES | (1) Advisory file locking; (2) All invalid metadata and all invalid file replicas are deleted when closing a file; (3) Read lock and a write lock are supported for the whole file or a region of a file; (4) Processes access the same file replica when the file is locked; (5) Client cache is disabled in the locked region; (6) Gfsd updates the metadata after closing the file and when the connection from a client is broken. |
| NETWORK FEATURES | (1) scp; (2) GridFTP; (3) SMB. |

**Figure 14 Gfarm Features' Weight Comparison**

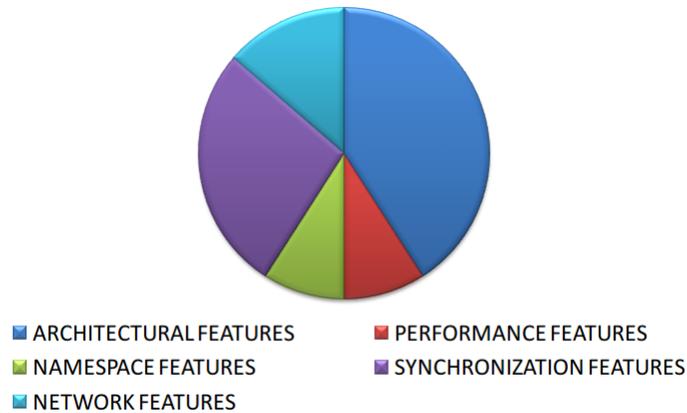

## 3.6 Parallel File Systems

### 3.6.1 PVFS

PVFS [33] , [34] , [35]  is the parallel file system for Linux cluster computing and has enabled low-cost clusters of high-performance PCs to address parallel applications with large-scale I/O needs. PVFS provides a clusterwide consistent name space, enables user-controlled striping of data across disks on different I/O nodes.

The PVFS was designed with the following goals in mind:
- It must provide high bandwidth for concurrent read/write operations from multiple processes or threads to a common file.
- It must support multiple APIs: a native PVFS API, the UNIX/POSIX I/O API, as well as other APIs such as MPI-IO.

- Common UNIX shell commands, such as ls, cp, and rm, must work with PVFS files.
- Applications developed with the UNIX I/O API must be able to access PVFS files without recompiling.
- It must be robust and scalable.
- It must be easy for others to install and use.

**Table 15 PVFS Features Classification**

| PVFS | |
|---|---|
| ARCHITECTURAL FEATURES | (1) Parallel file system; (2) Explicit state machine system; (3) File data is split into datafiles; (4) One type of server process (metadata OR data pvfs2-server); (5) UNIX files hold file data; (6) Berkeley DB database holds metadata; (7) pvfs2-client API provides access to file system's operations; (8) ROMIO MPI-IO API; (9) Client obtain configuration information about the file system during start-up; (10) I/O daemons; (11) File striping across I/O nodes; (12) Metadata manager daemon; (13) System interface (libpvfs2) abstract the task of communicating with many servers concurrently; (14) Management interface for administrators, fsck, performance monitoring; (15) VFS support for PVFS2 by Linux kernel driver; (16) File data and metadata store in files on existing local file systems. |
| PERFORMANCE FEATURES | (1) MPI datatypes as efficient structured data support; (2) Interfaces to inform file system about access pattern; (3) Metadata for different files to be placed on different servers; (4) Stateless servers and clients (no locking subsystem); (5) Threads + State machines + Completion notifications == Avoids serialization of independent operations; (6) Native support for asynchronous operations; (7) MPI-IO support not through a UNIX-like interface; (8) Many servers as multiple paths to data; (9) Caching of the directory hierarchy for a configurable duration; (10) PVFS files are striped across a set of I/O nodes; (11) "Partitioned-file interface". |
| HIGH-AVAILABILITY FEATURES | (1) File striping across I/O nodes; (2) Stateless system without locks. |
| NAMESPACE FEATURES | (1) Clusterwide consistent name space; (2) Datafiles' reference table; (3) File name is resolved into an opaque reference; (4) File handles broadcast; (5) Open file concept is absent; (6) Caching of the directory hierarchy for a configurable duration; (7) Directory, metafile, datafile, symbolic link are visible to users; (8) Handle space ranges; (9) File system ID. |
| SYNCHRONIZATION FEATURES | (1) Different semantics of coherency of the file system view; (2) PVFS2 does not provide POSIX semantics; (3) PVFS does not provide guarantee about the atomicity of read and write operation performed concurrently; (4) PVFS doesn't use a locking subsystem; (5) All of APIs are nonblocking; (6) Deleted file is removed immediately; (7) Nonconflicting writes semantic. |
| NETWORK FEATURES | (1) Buffered Messaging Interface (BMI); (2) Direct communication with I/O nodes after file open. |
| SECURITY FEATURES | (1) Two step permission checking (VFS + metadata server); (2) A client may lose the ability to access a file due to permission change. |

**Figure 15 PVFS Features' Weight Comparison**

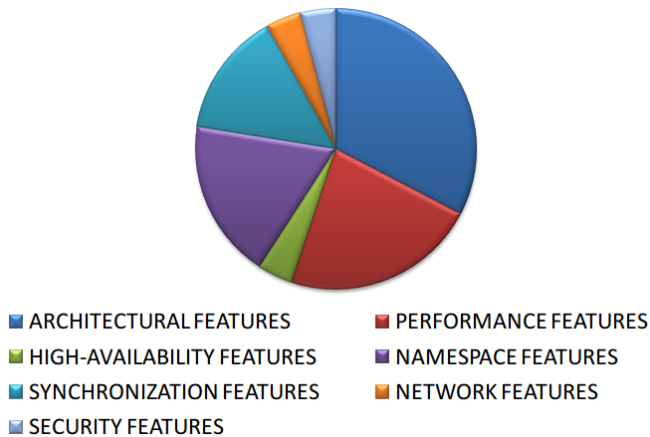

- ARCHITECTURAL FEATURES
- PERFORMANCE FEATURES
- HIGH-AVAILABILITY FEATURES
- NAMESPACE FEATURES
- SYNCHRONIZATION FEATURES
- NETWORK FEATURES
- SECURITY FEATURES

It is possible to see from the Figure 15 that the most significant classes of features in PVFS are: (1) Performance optimization features; (2) Namespace features; (3) Synchronization features.

### 3.6.2 Lustre

Lustre [36] , [37] , [38] , [38]  is a GNU General Public licensed, open-source distributed parallel filesystem. The name Lustre embodies "Linux" and "Cluster". Lustre focuses on scalability for use in large computer clusters, but can equally well serve smaller commercial environments through minor variations in the implementation and deployment of the modules that make up the system.

Lustre uses a modified version of the ext4 journaling file system to store data and metadata. This version, called ldiskfs, has been enhanced to improve performance and provide additional functionality needed by Lustre.

Lustre supports a variety of high performance, low latency networks and permits Remote Direct Memory Access (RDMA) for Infiniband (OFED). This enables multiple, bridging RDMA networks to use Lustre routing for maximum performance.

Lustre offers active/active failover using shared storage partitions for OSS targets (OSTs) and active/passive failover using a shared storage partition for the MDS target (MDT). This allows application transparent recovery. Lustre can work with a variety of high availability (HA) managers to allow automated failover and has no single point of failure (NSPF). Multiple mount protection (MMP) provides integrated protection from errors in highly-available systems that would otherwise cause file system corruption.

Any clients can operate on the same file and directory concurrently. A Lustre distributed lock manager (DLM) ensures that files are coherent between all the clients in a file system and the servers. Multiple clients can access the same files concurrently, and the DLM ensures that all the clients see consistent data at all times.

The distribution of files across OSTs can be configured on a per file, per directory, or per file system basis. This allows file I/O to be tuned to specific application requirements. Lustre uses RAID-0 striping and balances space usage across OSTs.

Lustre has a dedicated MPI ADIO layer that optimizes parallel I/O to match the underlying file system architecture.

**Table 16 Lustre Features Classification**

| Lustre | |
|---|---|
| ARCHITECTURAL FEATURES | (1) Cluster; (2) Management server (MGS); (3) Object-based filesystem; (4) Metadata servers (MDSs); (5) Object storage servers (OSSs); (6) Clients; (7) Object Storage Target (OST); (8) Standard POSIX I/O system calls; (9) Metadata Target (MDT); (10) RAID 0 pattern (data is "striped" across a certain number of objects). |
| PERFORMANCE FEATURES | (1) Modified version of the ext4 journaling file system; (2) Lustre supports mmap() file I/O; (3) Lustre uses RAID-0 striping and balances space usage across OSTs; (4) MPI ADIO layer that optimizes parallel I/O; (5) Ability to stripe data across multiple OSTs in a round-robin fashion. |
| RELIABILITY FEATURES | (1) Checksum of all data sent from the client to the OSS; (2) Distributed file system check (lfsck). |
| HIGH-AVAILABILITY FEATURES | (1) Active/active failover using shared storage partitions for OSS targets (OSTs); (2) Active/passive failover using a shared storage partition for the MDS target (MDT); (3) High availability (HA) manager; (4) Multiple mount protection (MMP) provides integrated protection from errors; (5) Availability is accomplished by replicating hardware and/or software; (6) A pair of servers with a shared resource. |
| NAMESPACE FEATURES | (1) Single, coherent, synchronized namespace; (2) POSIX-compliant filesystem; (3) Extended attribute (EA) describes the mapping between file object id and its corresponding OSTs; (4) Each filename points to an inode. The inode contains all of the file attributes. |
| SYNCHRONIZATION FEATURES | (1) In a cluster most operations are atomic; (2) Distributed lock manager (DLM); (3) Two types of request: lock related and data related. |
| NETWORK FEATURES | (1) Remote Direct Memory Access (RDMA) for Infiniband (OFED); (2) Re-exported using NFS or CIFS (via Samba); (3) All client/server communications in Lustre are coded as an RPC request and response; (4) Lustre Networking (LNET). |
| SECURITY FEATURES | (1) TCP connections only from privileged ports; (2) Group membership handling is server-based; (3) ACLs. |
| SCALABILITY FEATURES | (1) A new OSS with OSTs can be added to the cluster without interrupting any operations. |

**Figure 16 Lustre Features' Weight Comparison**

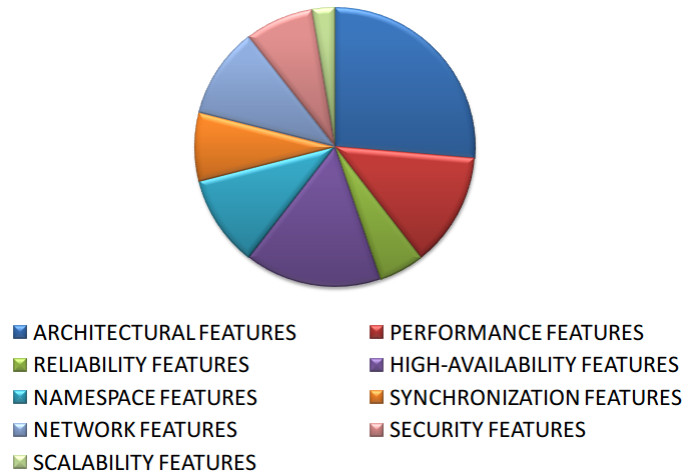

It is possible to see from the Figure 16 that the most significant classes of features in Lustre are: (1) High-availability features; (2) Performance optimization features; (3) Network features; (4) Namespace features.

### 3.6.3 GPFS

A GPFS [40] , [41] , [42] , [43]  file system is built from a collection of disks which contain the file system data and metadata. A file system can be built from a single disk or contain thousands of disks storing Petabytes of data. A GPFS cluster can contain up to 256 mounted file systems.

Applications can access files through standard UNIX file system interfaces or through enhanced interfaces available for parallel programs. Parallel and distributed applications can be scheduled on GPFS clusters to take advantage of the shared access architecture. Parallel applications can concurrently read or update a common file from multiple nodes in the cluster. GPFS maintains the coherency and consistency of the file system using a sophisticated byte level locking, token (lock) management and logging. In addition to standard interfaces GPFS provides a unique set of extended interfaces which can be used to provide high performance for applications with demanding data access patterns.

GPFS achieves high performance I/O by:
- Striping data across multiple disks attached to multiple nodes.
- High performance metadata (inode) scans.
- Supporting a large block size, configurable by the administrator, to fit I/O requirements.
- Utilizing advanced algorithms that improve read-ahead and write-behind file functions.
- Using block level locking based on a very sophisticated scalable token management system to provide data consistency while allowing multiple application nodes concurrent access to the files.

**Table 17 GPFS Features Classification**

| | GPFS |
|---|---|
| ARCHITECTURAL FEATURES | (1) Standard UNIX file system interface; (2) Enhanced interfaces available for parallel programs; (3) Data Management API (DMAPI); (4) Storage pool (groups of disks within a file system); (5) File placement policies; (6) mmfsd (persistent GPFS daemon); (7) Virtual Shared Disk (VSD) layer; (8) Multi-threaded architecture; (9) Metanode (Any updates to the inode information for the file are negotiated with the metanode). |
| PERFORMANCE FEATURES | (1) Set of extended interfaces for demanding data access patterns; (2) Striping data across multiple disks attached to multiple nodes; (3) High performance metadata (inode) scans; (4) Large block size; (5) Advanced algorithms of read-ahead and write-behind; (6) Block level locking based on scalable token management system; (7) Recognition and optimization I/O access for typical access patterns; (8) "Client-side cache" design; (9) GPFS implements striping in the file system; (10) Large files in GPFS are divided into equal sized blocks, and consecutive blocks are placed on different disks in a round-robin fashion; (11) GPFS prefetches data into its buffer pool issuing I/O requests in parallel; (12) Dirty data buffers that are no longer being accessed are written to disk in parallel; (13) Extensible hashing to organize directory entries within a directory. |
| RELIABILITY FEATURES | (1) Snapshot; (2) Journal or write-ahead log records all metadata updates on each node; (3) Dual-attached RAID controllers; (4) Replication. |
| HIGH-AVAILABILITY FEATURES | (1) Monitoring the health of the file system components and automatic recovery; (2) Journal logs, metadata and data replication; (3) Connection retries; (4) File striping; (5) Replacement of failed token manager or stripe group manager; (6) A quorum is required for successful file system mounts; (7) Extensive logging; (8) Heartbeat messages; (9) Process group membership protocol; (10) GPFS |
|  | allows accessing a file system only by the group containing a majority of the nodes in the cluster. |
| NAMESPACE FEATURES | (1) GPFS provides scalable metadata management by allowing all nodes of the cluster accessing the file system to perform file metadata operations; (2) GPFS manages metadata at the node which is using the file or in the case of parallel access to the file, at a dynamically selected node which is using the file; (3) Filesets provide an administrative boundary that can be used to set quotas and be specified in a user defined policy to control initial data placement or data migration. |
| SYNCHRONIZATION FEATURES | (1) Byte level locking; (2) Token (lock) management; (3) Distributed locking; (4) Various synchronization approaches; (5) Centralized global lock manager; (6) Data shipping mode; (7) Shared write lock on the inode; (8) Particular file's metanode election. |
| NETWORK FEATURES | (1) NFS or Samba. |
| SCALABILITY FEATURES | (1) Multiple nodes can act as token managers; (2) Allowing all nodes of the cluster accessing the file system to perform file metadata operations. |

**Figure 17 GPFS Features' Weight Comparison**

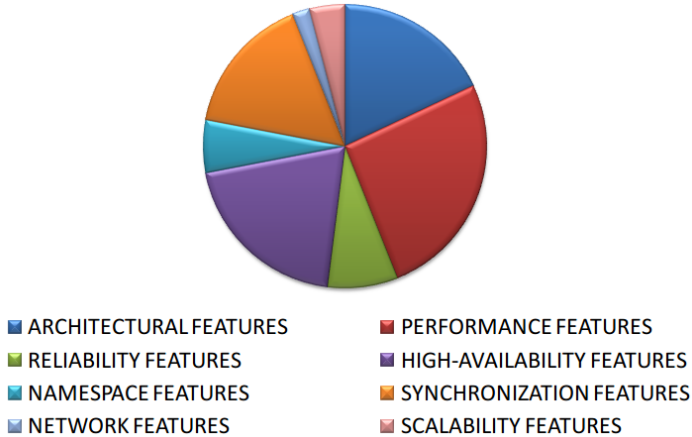

It is possible to see from the Figure 17 that the most significant classes of features in GPFS are: (1) Performance optimization features; (2) High-availability features; (3) Synchronization features.

## 3.7  Design Goals vs. Feature Classes

From the common point of view, a file system architect has a vision of design goals during elaboration of file system architecture. These goals are the basis for research of internal techniques that will be used for file system implementation. Thereby, as a result, design goals define importance of different classes of features for concrete file system architecture.

Table 18 is an attempt to compare design goals of analyzed file systems with the most weighty feature classes of its. It was extracted the most important design goals from concrete file systems' descriptions and it was chosen three the most important features classes.

**Table 18 Design Goals vs. Feature Classes**

| File System | Design Goals | Feature Class I | Feature Class II | Feature Class III |
|---|---|---|---|---|
| HDFS | (1) To be a file system designed for storing very large files with streaming data access patterns, running on clusters on commodity hardware; (2) The most efficient data processing pattern is a write-once, read-many-times pattern; (3) The time to read the whole dataset is more important than the latency in reading the first record. | Network features | Reliability features | High-availability features |
| GFS | (1) To be a distributed file system to be run on clusters up to thousands of machines; (2) To run on commodity hardware; (3) To be a high-available file system; (4) To be targeted at a particular set of usage scenarios is optimized for usage of large files only with space efficiency being of minor importance; (5) The majority of files can be considered as being append-only or even immutable (write once, read many). | Synchronization features | Reliability features | High-availability features |
| InterMezzo | (1) To be scalable and high-available file system. | Synchronization features | - | - |
| Coda | (1) To be a scalable, secure, and highly available distributed file system; (2) To achieve a high degree of naming and location transparency so that the system would appear to its users very similar to a pure local file system; (3) To be a highly-available file system with high degree of failure transparency. | High-availability features | Namespace features | Performance optimization features |
| Ceph | (1) To be scalable file system; (2) To utilize a highly adaptive distributed metadata cluster architecture; (3) To be a highly-available file system. | Performance optimization features | Namespace features | Reliability features |
| DDFS | (1) To support use cases that are typical for Disco and MapReduce in general: storage and processing of massive amounts of immutable data; (2) To be a tag-based filesystem; (3) To operate on commodity hardware. | Namespace features | High-availability features | Synchronization features |
| zFS | (1) To be a decentralized file system that distributes all aspects of file and storage management over a set of cooperating machines interconnected by a high-speed network; (2) To use the memory of all participating machines as a global cache to increase performance; (3) To achieve almost linear scalability: the addition of machines will lead to an almost linear increase in performance. | Namespace features | Performance optimization features | Security features |
| Zebra | (1) To provide a file transfer rate that scales with the number of servers; (2) To support UNIX workloads are characterized by short file lifetimes, sequential file accesses, infrequent write-sharing of a file by different clients, and many small files; (3) To provide file service despite the loss of any single machine in the system. | Performance optimization features | Reliability features | Synchronization features |
| PlasmaFS | (1) To be a distributed filesystem for large files, implemented in user space; (2) To provide data safety and clear query semantics by means of ACID transactions. | Namespace features | Performance optimization features | Reliability features |
| Xsan | (1) To be a high-performance storage area network (SAN) file system; (2) To provide the | Performance | Network features | High-availability |
| | highest level of data availability. | optimization features | | features |
| CXFS | (1) To be a shared XFS filesystem that allows groups of computers to coherently share large amounts of data while maintaining high performance; (2) To provide a single-system view of the filesystems. | Performance optimization features | High-availability features | Namespace features |
| ACFS | (1) To be scalable file system. | Performance optimization features | Reliability features | High-availability features |
| ExaFS | (1) To provide single global file name space; (2) To provide fully distributed file system; (3) To support scalability growth in both capacity and bandwidth independently; (4) To provide fully clustered solutions that allow high availability, reliability, and reduces downtime. | Performance optimization features | High-availability features | Reliability features |
| Gfarm | (1) To provide a Grid file system that integrates local disks of compute nodes in Computational Grid; (2) To provide transparent sharing of file data in a Grid; (3) To provide parallel and distributed computing associating Computational Grid and Data Grid. | Synchronization features | Network features | - |
| PVFS | (1) To provide high bandwidth for concurrent read/write operations from multiple processes or threads to a common file; (2) To support multiple APIs; (3) To be scalable. | Performance optimization features | Namespace features | Synchronization features |
| Lustre | (1) To be scalable for use in large computer clusters; (2) To support a variety of high performance, low latency networks; (3) To tune the file I/O to specific application requirements; (4) To optimize parallel I/O. | High-availability features | Performance optimization features | Network features |
| GPFS | (1) To provide standard UNIX file system interfaces for parallel programs; (2) To provide concurrent read or update a common file from multiple nodes in the cluster; (3) To provide byte level locking, token (lock) management and logging. | Performance optimization features | High-availability features | Synchronization features |

It is possible to extract from the Table 18 that the most important design goals are: **(1) To be a high-available file system**; **(2) To be scalable file system**; **(3) To provide special namespace features**; **(4) To be a high-performance file system**; **(5) To run on commodity hardware**. It can be concluded that such distribution of design goals by importance is reflected end-users' expectations. First of all, an end-user wants to have access to data "always" without presence of failure or denial of service. Secondly, it treated as very desirable feature such opportunity as potential easy scalability of system without presence of bottlenecks in services or performance. Thirdly, usually, end-user wants to have special namespace features that are optimized for target use-cases or workloads. Of course, the performance and efficiency is very important for end-users but with using cheap and commodity hardware. It is possible to see that these goals are contradictory to each other. As a result, every file system contains a set of trade-offs.

**Table 19 File System vs. Feature Class**

| File System vs. Feature Class | | |
|---|---|---|
| CXFS, ExaFS, GPFS | Performance optimization features | High-availability features |
| Zebra, ACFS | | Reliability features |
| Ceph, PVFS | | Namespace features |
| Xsan | | Network features |
| DDFS | Namespace features | High-availability features |
| zFS, PlasmaFS | | Performance optimization features |
| GFS | Synchronization features | Reliability features |
| InterMezzo, Gfarm | | Network features |
| Coda | High-availability features | Namespace features |
| Lustre | | Performance optimization features |
| HDFS | Network features | Reliability features |

However, Table 18 shows that the most important feature classes are: **(1) Performance optimization features**; **(2) Namespace features**; **(3) Synchronization features**; **(4) High-availability features**. It means that, first of all, file system architect takes in mind file system's efficiency. The performance is cornerstone of a file system architecture that plays definitive role for any designing or optimization attempts. Secondly, peculiarities of use-case or target workload can dictate necessity to use a special namespace paradigm. Thereby, specialized namespace API or file system objects' representation can be seen by a file system architect as a very promising and fundamental way of achieving design goals. Thirdly, the modern DFS lives in the heterogeneous environment of interacting nodes. And synchronization approaches are very important file system internal techniques. As a result, every performance or namespace approaches should be safety and efficient for the case of concurrent access to the file system's object. Finally, a high-availability is an important end-users' expectation. A modern DFS has to provide reliability techniques that guaranty a file system high-availability.

It is possible to see that it exist a difference between end-user expectations and a file system architect's vision of feature classes' importance. This difference is a basis for misunderstanding between an end-user and a file system architect. Moreover, it means that there isn't the perfect file system solution from an end-user point of view. An end-user takes in mind such set of requirements that can't be achieved without significant restrictions and trade-offs.

# 4 Special Class Features Analysis

## 4.1 Architectural Features

Architectural features characterize a file system design and describe the vision of principal architectural approaches that to define a file system's components and essence of internal interactions between of its. In other words, it is a fundamental file system concepts which define rules of play.

It is possible to analyze architectural features from the point of view several directions:

- **Specialized API**. This view tries to distinguish and to classify peculiarities of file system's APIs.
- **Pattern approaches**. This view tries to distinguish and classify a fundamental concepts which to define special peculiarities of target use-case or workload.
- **Server approaches**. This view introduces concepts of specialized nodes that realize different functions in file system architecture by means of interactions in the network.
- **Storage approaches**. This view tries to distinguish a special persistent storage oriented internal techniques.
- **Specialized network**. This view tries to distinguish network-oriented architectural solutions.

Such directions of architectural features analysis were chosen during preliminary trying of distribution features between sub-classes.

### 4.1.1 Architectural Features Analysis

Architectural features characterize the fundamental principles of organization and functioning the system at whole. Any file system architecture provides the opportunity to distinguish: (1) data access interface, (2) basic patterns of data access, (3) principles of organization of network infrastructure, and (4) principles of organization of subsystem of data storage.

**Table 20 Architectural Features' Subclasses**

| \multicolumn{3}{c|}{**Architectural Features**} |
|---|---|---|
| Specialized API | Not POSIX-compatible API | HDFS, GFS, DDFS, Ceph |
|  | POSIX-compatible API | Lustre, GPFS |
|  | Specialized Data Management API | PVFS, Lustre, GPFS |
| Pattern Approaches | Specialized Access Patterns | HDFS, GFS |
|  | Failure is a norm | HDFS, GFS, Ceph |
|  | Decouple data and metadata operations | HDFS, GFS, DDFS, Zebra, Ceph, PlasmaFS, zFS, Lustre, PVFS |
| Server Approaches | Single NameNode | HDFS, GFS, DDFS, Zebra |
|  | Multiple NameNodes | Ceph, PlasmaFS, Lustre |
|  | Specialized metadata server | PlasmaFS, XSAN, Gfarm, PVFS |
|  | DataNodes | HDFS, GFS, DDFS, Zebra, PlasmaFS, PVFS |
|  | Object Store Devices | Ceph, zFS, Lustre |

| Storage Approaches | Garbage Collection | DDFS, Zebra |
| --- | --- | --- |
| | Striping | Zebra, PVFS, Lustre, GPFS, ACFS |
| | Storage area network (SAN) | XSAN, CXFS |
| | RAID | XSAN, ExaFS |
| Disconnected Operations | Disconnected Operations | InterMezzo, Coda |
| Specialized network | Specialized network | ExaFS, CXFS |

**Table 21 Architectural Features vs. File Systems**

| | Architectural Features | | | | | |
| --- | --- | --- | --- | --- | --- | --- |
| | Specialized API | Pattern Approaches | Server Approaches | Storage Approaches | Disconnected Operations | Specialized Network |
| HDFS | Yes | Yes | Yes | - | - | - |
| GFS | Yes | Yes | Yes | - | - | - |
| InterMezzo | - | - | - | - | Yes | - |
| Coda | - | - | Yes | - | Yes | - |
| Ceph | Yes | Yes | Yes | - | - | - |
| DDFS | Yes | Yes | Yes | Yes | - | - |
| zFS | - | Yes | Yes | - | - | - |
| Zebra | - | Yes | Yes | Yes | - | - |
| PlasmaFS | - | Yes | Yes | - | - | - |
| XSAN | - | - | Yes | Yes | - | - |
| CXFS | - | - | - | Yes | - | Yes |
| ACFS | - | - | - | Yes | - | - |
| ExaFS | - | - | - | Yes | - | Yes |
| Gfarm | - | - | Yes | - | - | - |
| PVFS | Yes | Yes | Yes | Yes | - | - |
| Lustre | Yes | Yes | Yes | Yes | - | - |
| GPFS | Yes | - | - | Yes | - | - |

**Figure 18 Architectural Approaches' Weight Comparison**

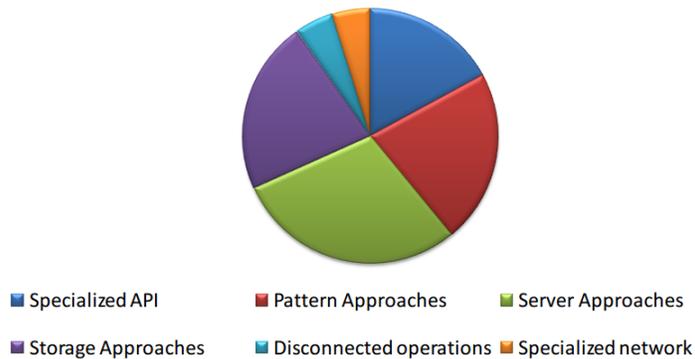

## 4.2 Performance Optimization Features

### 4.2.1 Performance Optimization Features Analysis

**Table 22 Performance Optimization Features' Subclasses**

| \multicolumn{3}{c}{Performance Optimization Features} | | |
|---|---|---|
| Specialized API | MPI-IO | PVFS, Lustre |
| | Specialized interfaces for data access patterns | PVFS, GPFS |
| Metadata Approaches | Knowledge of metadata popularity | Ceph |
| | Hashing to organize directory entries within a directory | Ceph, GPFS |
| | Special addressing scheme | Zebra, PlasmaFS |
| | Optimized metadata algorithms | CXFS, ACFS, GPFS |
| | Metadata load balancing | PVFS |
| Caching Approaches | Transactions batching | HDFS, GFS, Zebra |
| | File data caching | GFS, Coda, InterMezzo, Zebra, CXFS, ExaFS, GPFS |
| | Metadata caching | HDFS, GFS, Coda, Ceph, DDFS, Zebra, CXFS, PVFS |
| | Write operations caching | Coda, InterMezzo, Zebra, CXFS, ExaFS, GPFS |
| | Metadata access scenarios optimization | Ceph |
| | Lazily flushed journals strategy | Ceph |
| | Coherent cache (cooperative cache) | zFS, CXFS, Coda |
| | Preallocation | ACFS |
| | I/O parallelism (GPFS prefetches data into its buffer pool issuing I/O requests in parallel) | GPFS |
| Multi-Threading Approaches | Multiple file managers | zFS, Zebra, ExaFS |
| | Multi-threading | Coda, PlasmaFS, PVFS |
| | Idle Resource Utilization | ExaFS |
| | Stateless servers and clients (no locking subsystem) | PVFS |
| Storage Approaches | Specially optimized filesystem | Ceph, Lustre, Zebra |
| | Specially optimized low-level disk scheduler | Ceph |
| | Striping | Zebra, Lustre, GPFS, PVFS, ACFS |
| | Storage area network (SAN) | XSAN, CXFS, ACFS |
| | I/O parallelism | ACFS, GPFS, PlasmaFS, Coda |
| | Special I/O modes | XSAN, CXFS |
| | Load balancing | ACFS, XSAN |
| | Block/Fragment size | GPFS, CXFS, ACFS |
| | Access patterns optimization | Ceph, XSAN, GPFS |

| Network Approaches | Hardware optimizations | Coda, XSAN, CXFS, ACFS, ExaFS |
| --- | --- | --- |
| | Scheduling a task to the data location | HDFS |
| | Specialized protocol | GFS, CXFS |
| | Chunk size | GFS |
| | Server replication | Coda |
| | Caching | Coda, Zebra, zFS |
| | Specialized replication scheme | Ceph |
| | Multiple metadata servers | Ceph, CXFS |
| | Specialized network | CXFS, ExaFS |
| | Multipathing | Zebra, PVFS |
| | Load Balancing | ExaFS, DDFS, Ceph |

**Table 23 Performance Optimization Features vs. File Systems**

| | Performance Optimization Features | | | | | |
| --- | --- | --- | --- | --- | --- | --- |
| | Specialized API | Metadata Approaches | Caching Approaches | Multi-Threading Approaches | Storage Approaches | Network Approaches |
| HDFS | - | - | Yes | | - | Yes |
| GFS | - | - | Yes | | - | Yes |
| InterMezzo | - | - | Yes | | - | - |
| Coda | - | - | Yes | Yes | Yes | Yes |
| Ceph | - | Yes | Yes | | Yes | Yes |
| DDFS | - | - | Yes | | - | Yes |
| zFS | - | - | Yes | Yes | - | Yes |
| Zebra | - | Yes | Yes | Yes | Yes | Yes |
| PlasmaFS | - | Yes | | Yes | Yes | - |
| XSAN | - | - | | | Yes | - |
| CXFS | - | Yes | Yes | | Yes | Yes |
| ACFS | - | Yes | Yes | | Yes | - |
| ExaFS | - | - | Yes | Yes | Yes | Yes |
| Gfarm | - | - | | | - | - |
| PVFS | Yes | Yes | Yes | Yes | Yes | Yes |
| Lustre | Yes | - | | | Yes | - |
| GPFS | Yes | Yes | Yes | | Yes | - |

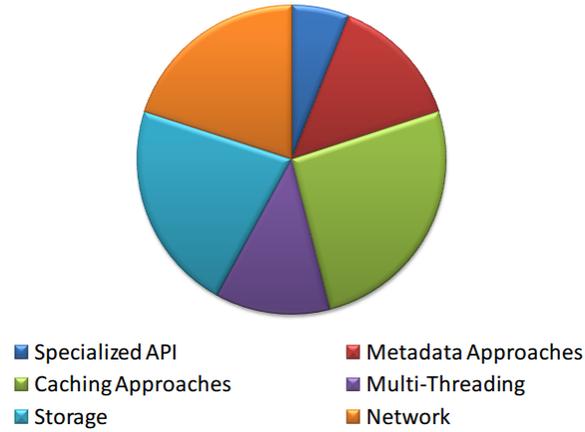

Figure 19 Performance Optimization Fetures' Weight Comparison

## 4.3 Reliability Features

### 4.3.1 Reliability Features Analysis

**Table 24 Reliability Features' Subclasses**

| \multicolumn{3}{c}{**Reliability Features**} |||
|---|---|---|
| Hardware Approaches | RAID | GPFS, ExaFS, CXFS |
| | Dual access | ExaFS |
| | Failover | ExaFS, Ceph |
| | UPS | ExaFS |
| | MultiSAN | XSAN |
| | Self-report | Ceph |
| | Active monitoring | Ceph |
| | Recoverable Virtual Memory (RVM) | Coda |
| Storage Approaches | Snapshot | GPFS, ACFS, GFS, HDFS |
| | Journaling approach | ExaFS, Ceph |
| | Mirroring | ACFS |
| | Transactions | PlasmaFS |
| | Replacement blocks for the file region allocation | PlasmaFS |
| | Cache entire files | Coda |
| | Checksumming | GFS, HDFS |
| Filesystem Approaches | Replication | GPFS |
| | Logging | CXFS, Zebra, GFS |
| | Checkpoint | HDFS |

| Network Approaches | Journaling | GPFS, ACFS, CXFS, XSAN, DDFS, GFS, HDFS |
| --- | --- | --- |
| | Checksumming | ACFS |
| | Data Transfer Protection | Zebra, Ceph, Lustre |
| | Specialized network | ExaFS |
| | Replication Approaches | ACFS, Coda, GFS, HDFS, ExaFS, Zebra |
| | Synchronous store operations | Zebra |
| | Logging | Zebra |
| | Distributed transactions | zFS |
| | Failover by means of quick rescan of the journal by another node | Ceph |
| | Disconnected operations | Coda |

**Table 25 Reliability Features vs. File Systems**

| | Reliability Features | | | |
| --- | --- | --- | --- | --- |
| | Hardware Approaches | Storage Approaches | Filesystem Approaches | Network Approaches |
| HDFS | - | Yes | Yes | Yes |
| GFS | - | Yes | Yes | Yes |
| InterMezzo | - | - | - | - |
| Coda | Yes | Yes | - | Yes |
| Ceph | Yes | Yes | - | Yes |
| DDFS | - | - | Yes | - |
| zFS | - | - | - | Yes |
| Zebra | - | - | Yes | Yes |
| PlasmaFS | - | Yes | - | - |
| XSAN | Yes | - | Yes | - |
| CXFS | Yes | - | Yes | - |
| ACFS | - | Yes | Yes | Yes |
| ExaFS | Yes | Yes | - | Yes |
| Gfarm | - | - | - | - |
| PVFS | - | - | - | - |
| Lustre | - | - | - | Yes |
| GPFS | Yes | Yes | Yes | - |

**Figure 20 Reliability Features' Weight Comparison**

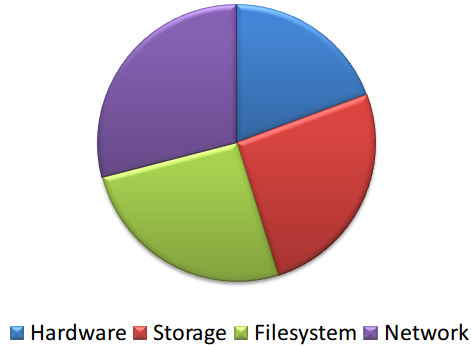

## 4.4 High-Availability Features

### 4.4.1 High-Availability Features Analysis

**Table 26 High-Availability Features' Subclasses**

| High-Availability Features | | |
|---|---|---|
| Hardware Approaches | Multipathing | XSAN, CXFS, ACFS |
| | Components redundancy | ExaFS, Lustre |
| | Dedicated network | CXFS, ExaFS |
| Network Approaches | Snapshot/Recovery | HDFS, GFS, ACFS, GPFS, Zebra, ExaFS |
| | Replication Strategy | HDFS, GFS, Coda, Ceph, DDFS, zFS, PlasmaFS, GPFS, PVFS |
| | Load Balancing | HDFS, GFS, XSAN, ACFS, ExaFS, DDFS |
| | Elections | Ceph, CXFS, GPFS |
| | Nodes Monitoring | GFS, CXFS, GPFS, Ceph, DDFS, ExaFS |
| | Failover | HDFS, GFS, XSAN, CXFS, ACFS, ExaFS, Lustre, GPFS |
| | Disconnected operations | Coda |
| | Garbage collection | DDFS, Zebra |

**Table 27 High-Availability Features vs. File Systems**

| | High-Availability Features | |
|---|---|---|
| | Hardware Approaches | Network Approaches |
| HDFS | - | Yes |
| GFS | - | Yes |
| InterMezzo | - | - |
| Coda | - | Yes |
| Ceph | - | Yes |
| DDFS | - | Yes |
| zFS | - | Yes |
| Zebra | - | Yes |
| PlasmaFS | - | Yes |
| XSAN | Yes | Yes |
| CXFS | Yes | Yes |
| ACFS | Yes | Yes |
| ExaFS | Yes | Yes |
| Gfarm | - | |
| PVFS | - | Yes |
| Lustre | Yes | Yes |
| GPFS | - | Yes |

Figure 21 High-Availability Features' Weight Comparison

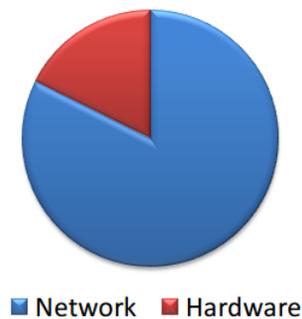

## 4.5 Synchronization Features

### 4.5.1 Synchronization Features Analysis

Synchronization of concurrent access to shared data is a key feature of any file system. Design and implementation of synchronization mechanisms is very complex and time-consuming task. Inefficient synchronization mechanisms can decrease file system performance significantly, whereas implementation phase's mistakes can be resulted in very sophisticated issues and failures in a file system's functioning.

Table 28 Synchronization Features' Classes

| Synchronization Features | | |
|---|---|---|
| Synchronization Mechanisms | Lock | GFS, Ceph, XSAN, CXFS, Gfarm, GPFS |
| | Lease | HDFS, GFS, zFS |

|  | Atomic mutation | GFS |
|---|---|---|
|  | Token mechanism | DDFS, CXFS, GPFS |
|  | Ticket system | PlasmaFS |
|  | Synchronization after close | InterMezzo, HDFS |
|  | Capabilities | Ceph |
|  | Versioning scheme of file updates | InterMezzo, Coda |
| Application-Level Mechanisms | O_LAZY flag | Ceph |
|  | Checkpointing | GFS |
|  | Data shipping mode | GPFS |
| Data Processing Patterns | Write-Once, Read-Many-Times Pattern | HDFS |
|  | Single Writer – Many Readers | HDFS |
|  | Producer-Consumer queue | GFS |
| Replica Processing | Stale replica processing | GFS |
|  | Corrupted/Lost chunk processing | GFS |
| Notification Mechanisms | Update notification | InterMezzo, Coda, Zebra |
| Disconnected operations | Disconnected operations | Coda |
| Transaction Approach | Transactions | Coda, DDFS |
|  | Cryptographic scheme | PlasmaFS |
| Request Approach | Replication request | InterMezzo |
|  | Cached object validation | InterMezzo |
|  | Reintegrate request | InterMezzo |
| No Locks | No metadata locks or leases | Ceph |
|  | Logging | InterMezzo, Zebra |
|  | Virtual stripe | Zebra |
|  | Replacement blocks for the file region allocation | PlasmaFS |

Known approaches of synchronization of concurrent access to shared data can be distributed between several classes:

- **Synchronization Mechanisms** are approaches of using synchronization primitives for concurrent access of several threads to shared data.

- **Application-Level Mechanisms** are approaches that provide to applications opportunity to define consistency of information that is modified concurrently from several threads.

- **Data Processing Patterns** are suggestion of fundamental data processing model that is a specialized file system's approach. Such approach restricts opportunity to concurrent data modification from several threads.

- **Replica Processing Approaches** are synchronization techniques of replicas of the same data in the case of data modification on one of nodes.

- **Notification Mechanisms** are techniques of notification of clients or data nodes about event of data modification on any file client or data node.

- **Disconnected operations** is approach in that temporary partitioning of file system on isolated fragments treats as normal situation. It takes place reintegration of data after recovery of connection between partitions.
- **Transaction Approaches** are techniques in that any file system operation is treated as indivisible transaction. Thereby, file system operation is meant as successful only after successful transaction commit.
- **Request Approaches** are techniques of modification of shared data by means requests to a file server.
- **No Locks Approaches** are mechanisms of concurrent modification of shared data by several threads without synchronization primitives.

It is possible to distribute all known approaches between two principal alternatives. One fundamental group of classes is realization of synchronization of concurrent access to shared data by means of: A. Synchronization primitives (Synchronization Mechanisms, Replica Processing Approaches); B. Restriction techniques (Application-Level Mechanisms, Transaction Approaches); C. Interaction by means of file server (Notification Mechanisms, Request Approaches). Second and opposite group of classes are attempts to realize concurrent multi-threaded data processing without using of synchronization primitives (Data Processing Patterns, Disconnected operations, No Lock Approaches).

There are such synchronization mechanisms for the case of using any of synchronization primitives:
1. Specialized synchronization primitives or synchronization objects on file system level.
2. Transaction approach of data modification.
3. Synchronization of access on application level.
4. Specialized file server which guarantees consistency of concurrently modified data.

**Table 29 Synchronization Features vs. File Systems**

| | Synchronization Features | | | | | | | | |
|---|---|---|---|---|---|---|---|---|---|
| | Synchronization Mechanisms | Application-Level Mechanisms | Data Processing Pattern | Replica Processing | Notification Mechanisms | Disconnected operations | Transaction Approach | Request Approach | No Locks |
| HDFS | Yes | - | Yes | - | - | - | - | - | - |
| GFS | Yes | Yes | Yes | Yes | - | - | - | - | - |
| InterMezzo | Yes | - | - | - | Yes | - | - | Yes | Yes |
| Coda | Yes | - | - | - | Yes | Yes | Yes | - | - |
| Ceph | Yes | Yes | - | - | - | - | - | - | Yes |
| DDFS | Yes | - | - | - | - | - | Yes | - | - |
| zFS | Yes | - | - | - | - | - | - | - | - |
| Zebra | - | - | - | - | Yes | - | - | - | Yes |
| PlasmaFS | Yes | - | - | - | - | - | Yes | - | Yes |
| XSAN | Yes | - | - | - | - | - | - | - | - |
| CXFS | Yes | - | - | - | - | - | - | - | - |
| ACFS | - | - | - | - | - | - | - | - | - |
| ExaFS | - | - | - | - | - | - | - | - | - |
| Gfarm | Yes | - | - | - | - | - | - | - | - |
| PVFS | - | - | - | - | - | - | - | - | - |
| Lustre | - | - | - | - | - | - | - | - | - |
| GPFS | Yes | Yes | - | - | - | - | - | - | - |

Table 29 shows that most of file systems realize synchronization of concurrently modified data on the basis of synchronization primitives. Another synchronization approaches are treated as auxiliary mechanisms that are efficient only in restricted use-cases. It is possible to say that all suggested alternatives to synchronization primitives can't be treated as equivalent solutions. However, evolution of HPC and distributed data processing is desperate in evolution of synchronization mechanisms. Such requirements of more efficient synchronization techniques stimulate elaboration of new approaches of concurrent access to shared data in distributed file systems. It is possible to distinguish such interesting and promising approaches as Data Processing Patterns, Disconnected operations, No Lock Approaches.

**Figure 22 Synchronization Features' Weight Comparison**

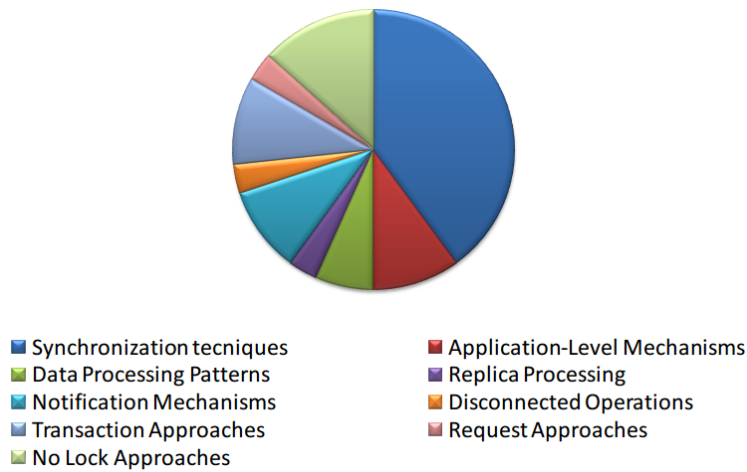

- Synchronization tecniques
- Application-Level Mechanisms
- Data Processing Patterns
- Replica Processing
- Notification Mechanisms
- Disconnected Operations
- Transaction Approaches
- Request Approaches
- No Lock Approaches

First of all, it is possible to point out that the whole weight of synchronization primitives is pretty high. It means that it has realized the necessity to elaborate some new approaches of more efficient synchronized access to shared data for specialized environments. However, the fundamental paradigm of using the synchronization primitives is treated like inevitable techniques by file system architects. Finally, it is possible to say that nobody suggested really good technique of synchronized access to shared data.

## 5   Comparative Features Analysis

It was used such methodology of concrete file system's features analysis. First of all, it was distinguished file system's features. Secondly, every feature was classified as contained by some feature class. As a result, such classification gives opportunity for comparison of features of different file systems inside a feature class.

But, also, it is possible to count for every feature class number of features inside a class for all file systems. Such calculation gives opportunity to construct a diagram with comparison of weight of feature classes (see Figure 23).

Figure 23 Feature Classes' Weight Comparison

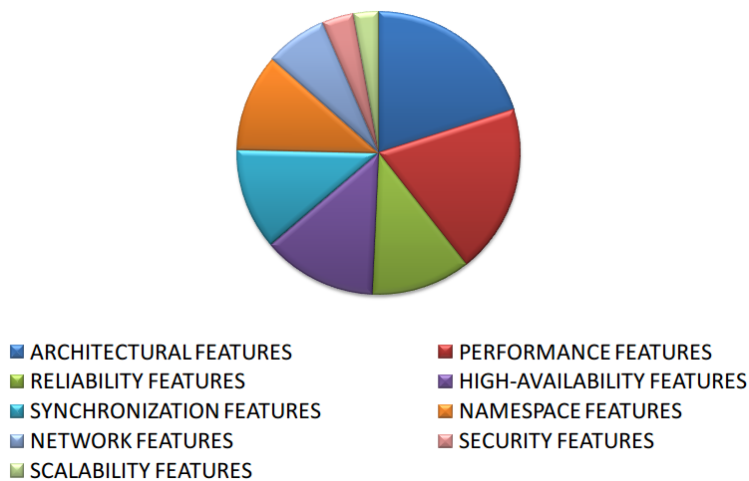

It can be concluded that the performance optimization features have the greatest weight with comparison of other feature classes. However, it can be stated that end-users don't satisfied by current file systems' performance. It means that needs in data processing are evolving faster than file systems evolution in whole. And an end-user hasn't breakthrough in data processing performance in spite of progress in data storage hardware area. The complexity of file system software stack is a reason of significant overhead that takes place during access and storing of persistent data. Big data use-case adds complexity because of necessity of distributed data processing on many nodes. Such situation gives basis for conclusion that efforts in only file systems' performance improvement can't resolve the problem in whole. It is more promising way to search new paradigms of representation and processing of data.

Significant importance in a modern technological world has safe and controllable access to user information that file systems store. However, it can be concluded that file systems' security features are developed not as extensively as it would. Needs in safe and controllable access to user data are growing with data capacity growing. But access control mechanisms decrease a file system performance by nature and, as a result, it worsens performance of data processing in whole. Moreover, file system software stack has different vulnerabilities that give opportunities for exploiting it as backdoors for system attacks with the purpose of file system's denial of service achieving. Thereby, it has to be stated a necessity in deeper attention to the problem of safe and controllable access to file systems.

File systems' scalability features also haven't enough progress in the environment of continuous data capacity growing and enhancing complexity of requests of data retrieval. Scalability has complex nature because of impossibility to have algorithms that are efficient for any capacity of input data. Factually, an end-user has necessity in file system solutions that are equally efficient for any size of data capacity and file sizes. From one viewpoint, every file system solution should be transparently scalable. But, really, not every solution can be as scalable as efficient. So, from other viewpoint, it can be used approach of compilation of several solutions in one complex solution. Every concrete solution will be efficient for some range of data capacity. The approach of using complex solution can require specialized API with the purpose to give a hint to file system about most efficient mechanism in current workload.

Networking plays key role in performance, reliability and high-availability features of DFS. Namely, networking significantly complicates synchronization of file system operations and, as a result, can decrease performance in some environment. From one viewpoint, networking influences on every DFS approach. But, from another viewpoint, key paradigms of file system technology have not distributed nature that conflicts with nature of networking technology. Factually, all complexity of DFS technology arises from this contradiction. As a result, it can be stated necessity of revisiting as file system technology's paradigms as network access paradigm with the purpose of elaboration of a new vision of distributed data access and processing.

Analysis of key goals of different DFS results in distinction of key expectations of an end-user. It is possible to state that an end-user expects: (1) high-available file system; (2) scalable file system; (3) file system with specialized paradigm of representation and storing data that oriented on user data peculiarities; (4) high-performance file system. Factually, it can be concluded a necessity to spend more efforts in the direction of elaboration scalability approaches of DFS in the environment of data capacity growing. The problem of distributed file system's high-availability can be treated as sub-problem of scalability problem. Growing data capacity can be resulted in file system performance degradation and situations of denial of service. There are efforts in DFS area with suggestion of specialized paradigm of representation and storing data. Moreover, file system API can be modified also. Thereby, such efforts can be a basis for elaboration of a new vision of paradigm of distributed storing and processing of data.

# 6  DFS Optimization and Design Vision

## 6.1  What Society and Business Does Expect from File Systems?

Data capacity is growing exponentially with every year. It makes stricter such requirements of file systems' architecture and internal techniques as **ability to store and provide access to big data without performance bottlenecks**. Performance and data capacity are related features because file size defines efficiency of metadata operations very frequently. Big data is a source of new complex problems in the field of data processing and representation of results. There is a wide area of problems in the field of big data processing that require more efficient file systems' internal techniques, new algorithms and paradigms of parallel data processing. Efficiency and profitability of business processes is defined by opportunities to analyze big data and to make business decisions in the real time.

The problem of big data analysis and processing is very complex and to require more relaxed file systems' API, especially for the case of distributed and parallel data processing. There is a tendency to relax or to expand a file system's API for the more efficient support of distributed data processing paradigm, peculiarities of processing data or used algorithms, and specialized disciples of replication and/or migration of data. New physical principles of persistent data storage (for example, phase change memory) raise very complex problems for file system architects. From one viewpoint, such technologies hide a great opportunity for enhancement of file systems' efficiency but, from another viewpoint, it raises a question about necessity of file systems as the technological approach at whole. **Evolving nature of information and approaches of data processing requires changing as architecture approaches as API approaches of future file systems**.

The reliability and high-availability are very important features of DFS. Opportunity to have access to data "always" is a crucial for an end-user. Denial of file system's service can be a reason of financial losses, data losses and dangerous situation. Modern evolution of data processing, evolving importance of information systems for social and business organizations needs in new requirements of reliability and high-availability of data storing and accessing. Information culture of modern society results in the expectation of an end-user **to have reliable and highly-available file system with possibility accessing to data even in the case of hardware failure**.

Nowadays, data reliability includes expectation of an end-user **to protect data against of sudden information losses, malicious activity and theft**. The reason of data losses can be a different. It can be result of sudden data deletion/corruption by means of wrong user's actions, erroneous applications working and bugs in file system driver. Malicious data corruption can take place because of virus activity or DoS-attacks on the system. Now information is a key asset of a company. Thereby, phenomenon of theft of commercial and technological information is a very important. All above-mentioned issues have to be solved by file system's approaches without performance bottlenecks.

Mobility is a key feature of advanced technological solutions. Modern user has heterogeneous information environment is built on the basis of different gadgets and computer devices. Such complex environment requires in new standards of file system's services. First of all, it means of **file system availability by means of variety of widely used access protocols**. More important requirement can be an opportunity **to process data by means of disconnected operations**. And, finally, it is possible to add a requirement of **transparent migration of data between user's gadgets**. Thereby, providing of a mobile data access is a key requirement of architecture of future file systems. And such goal is a challenging task for file system architect.

Growing data capacity makes the problem of data analysis and data processing more difficult and complex. The data has any value only in the case of opportunity to do analytical conclusions during reasonable duration of data processing. If we have to process data during a time that is greater than available time for making a business decision, then, such data hasn't any value. However, nowadays, business processes give rise to more and more information that needs to be analyzed. As a result, **an end-user expects new file system technologies for the environment of growing data capacity. Such technologies should provide big data analysis and transformation algorithms that guarantee making an important business decisions in time**.

Searching and sorting algorithms were always the key items of paradigm of storing digital information. Existing file systems' paradigms of searching and sorting information don't satisfy needs of an end-user in the environment of exponential growing of information. Moreover, the classical file system's paradigm of folders' hierarchy can be a reason of low performance of file system, duplication of data in files and can require complex searching algorithms in the case of big data. It can slow down processing of data and to deepen the big data problem. Thereby, an end-user expects **efficient file system's paradigm of big data sorting that can be a foundation for fast searching algorithms**.

The key point of user interaction with information is getting of data and representation of it by means of UI. The big data deepens this problem because it requires new approaches and more efficient algorithms of data access. Also, there is a contradiction between freedom of access to data and necessity to grant access to data on the basis of access rights. The big data widen a distance between a user and searching data with every day. Growing data capacity requires new approaches of file system architecture and more efficient algorithms of analysis and retrieval of searching data. Moreover, there are such data capacity and retrieval requests that require specialized high-performance hardware and inadequate calculation time. Thereby, **an end-user expects simple UI of fast getting of searching information. And it is expected a time of the search that correlates with lifetime of a task of data analysis in the environment of exponentially growing big data**.

Representation of digital data is a complex problem. Growing data requires new approaches of information ordering. A good approach of ordering can provide as faster data retrieving as faster analysis and transformation of information. Moreover, such approaches can be a basis for more efficient UI that can improve user perception of big data and to provide a new paradigm of management and transformation of data. Factually, **an end-user expects solving of the problem of big data complexity by means of new approaches of information ordering and UI paradigm**.

Usually, file system keeps data in the form of files. And file is a binary stream from the file system's viewpoint. But every file contains as metadata as user data that ordered with coincidence of some format. Retrieval of user data from file can be made by specialized application is identified by signature (and file extension). Factually, file is an essence which isolates user data by means of binary boundary. User data needs in ordering but existing paradigm of information ordering in modern file systems contradicts to the nature of user data. An end-user is a data-centric and he needs in ordering of information itself. But modern file systems represent data in the form of files tree that deepens complexity of data analysis very frequently. **An end-user expects "integrated" information space which can provide "natural" (task-oriented) and self-optimizing user data ordering without any dependence from internal format of data in file system**.

It is possible to distinguish several levels of abstraction: (1) Abstraction level of physical storage that encapsulates peculiarities of storing, reading and writing data; (2) File system abstraction level that encapsulates volume metadata format and internal file system's techniques; (3) Abstraction level of file that encapsulates data format and retrieval/writing mechanisms. Modern file formats and scenarios of data access very frequently contradict with physical storage's efficient technique of data access. Thereby, it is possible to indicate conflict of abstraction levels. As a result, physical storage can't be used over physical limits of efficiency. **An end-user expects using of physical data storage over physical limits of efficiency**.

Nowadays, situation of storing a huge capacity of data in concrete file system can be a problem. The key peculiarity of this problem can define inexpediency of changing technology of data storing. Necessity to copy or to move a huge capacity of data can be inadmissible. Big data deepens this issue because of capacity of data and fast evolution of distributed file systems. Thereby, a new DFS technologies' evolution can be suppressed by necessity to migrate a huge data capacity from old file system to a new one. Factually, file system architects have to take into consideration this technological issue. And it can influence on architecture of future file systems. As a result, **an end-user expects transparent integration of stored user data in the environment of new distributed file systems without necessity to move data between file systems**.

Modern business processes require in scalability of file systems because of necessity to process growing capacity of data. Nowadays, **an end-user expects ease file system's scalability**

**without any performance bottlenecks or denial of service. Moreover, file system should provide as quantitative as qualitative scalability features**.

File system's cost of service (COS) plays key role for the choosing of data warehouse architecture. Frequently, it needs to use specialized hardware solutions for achieving goal of efficient, reliable and high-available data storage and processing. As a result, COS of such solutions can be very expensive. Thereby, an end-user treats solutions on the basis of commodity hardware as a reasonable from the COS viewpoint. Finally, it is possible to conclude that **an end-user chooses a file system provides reasonable combination of COS and functionality**.

## 6.2  File System Optimization Criticism

A file system optimization is a most common task in the file system field. Usually, it is seen as the key file system problem. Moreover, it is possible to state that optimization is dominant in commercial development. A problem of a new file system architecture development arises more frequently in academia.

A problem definition of file system optimization can be defined by customer of a project as not optimization problem but as desire to achieve a file system's performance enhancement at several times. A customer can treat file system's performance as value that guarantees achievement by end-user the improvement of the whole system performance and functional opportunities in all possible use-cases without getting into account the workloads' peculiarities.

What is a file system performance? The problem of a file system performance definition and measurement is complex. One of the definitions of performance is: A performance is an amount of operations to the time used. It is possible to define a file system performance as an amount of I/O operations to the time used. But, however, many factors can influence on amount of I/O operations. It can be: (1) state of the volume; (2) file system internal techniques; (3) complexity of file system software stack; (4) use-case or workload type. The duration of file system operations can be defined also by: (1) algorithms of I/O scheduler and task scheduler; (2) the whole system's utilization. From other viewpoint, measurement procedure also can influence on measured value of file system performance. Moreover, internal file system's techniques can initiate I/O operations in background (for example, GC activity). Factually, it needs to take into consideration only attempts of file system performance measurement in restricted and reproducible environment. File system performance is slightly abstract notion that not define directly the whole computer system performance. But file system operations are a part of the whole overhead that computer system has. And namely the whole computer system performance can be seen by end-user. Wrong or inefficient using of file system API by applications is very frequently a reason of inefficiency a computer system at whole. However, evolving workloads makes file systems really inefficient in some use-cases. The reasons of new complexity of workloads are: (1) complicating nature of data processing; (2) increasing of data capacity; (3) complication of software stack; (4) tendency of evolution into SMP and heterogeneous systems.

An end-user can treat file system performance as a key problem of file system evolving as technology. Such understanding arises from common treatment of persistent memory as slow subsystem. As a result, problem of improving performance of data processing treats as a problem of file system performance optimization. However, evolution of physical technologies of persistent data storage requires significant changing of conceptions and approaches of file systems' internal techniques. Factually, only trying to improve file system efficiency can't resolve all issue of file systems as technological direction. Moreover, it can impede evolution of file system technology at whole. It is impossible to satisfy end-user's expectations by means of file systems optimization only. New persistent storage technologies can question about file systems necessity at whole without suggestion of revolutionary new file system's approaches. However, file system contains paradigm of information structuring that is very important for an end-user as a human being. But this paradigm has to be evolved with the purpose of suggestion of new (more intuitive) approaches of access to self-organizing information. From other view, existing digital information should be accessible to an end-user in usual way.

It needs to distinguish the two classes of tasks: (1) optimization task; (2) task of elaboration a new architecture vision or paradigm. But, frequently, project goal degenerates into optimization task which is meant really elaboration of a new paradigm. What does file system optimization task is? File system optimization task is resolving of bottleneck for concrete environment or/and workload. Factually, file system optimization task is searching of a best solution on the basis of taking into account peculiarities of hardware, workload or algorithm. As a result, optimization task means modification of existing software code, algorithm or approach.

But in what type of problems can be identified an end-user's expectations? End-user expectations are complex and contradictory set of requirements. Only optimization tasks can't resolve all current needs of an end-user in file system field. End-user's expectations require resolving tasks of a new architecture vision or paradigm elaboration.

## 6.3 AverageDFS

Suppose that it needs to elaborate a vision of a new DFS architecture and internal techniques of it. It is possible to take into account only such internal techniques that have the largest weight in a feature class. Such vision can be useful for analysis of internal techniques of existing file system or for elaboration of preliminary vision of new file system architecture. Factually, if any approach is used in several file systems then it means that this approach is a typical known technical solution.

It is possible to distinguish internal techniques with the largest weight in a feature class on the basis of made file systems' classification and comparative analysis. Thereby, on the basis of such approach it is made description of internal techniques' set for an abstract average DFS (AverageDFS).

**Table 30 Architectural Features of AverageDFS**

| AverageDFS: Architectural Features ||
|---|---|
| Pattern approach | Decouple data and metadata operations |
| Server approach | DataNodes |
| | NameNodes |
| | Specialized metadata server |
| Storage approach | Striping |

The AverageDFS's architecture has pattern, server and storage approaches in the basis. First of all, it declares "decouple data and metadata operations" approach. Secondly, it suggests to have cluster is contained as many DataNodes as many NameNodes. Moreover, metadata service can operate on the basis of specialized server technology (LDAP, database and so on). And, finally, striping between DataNodes can be a base storage technology.

**Table 31 Performance Optimization Features of AverageDFS**

| AverageDFS: Performance Optimization Features ||
|---|---|
| Caching approach | File data caching |
| | Metadata caching |
| | Write operations caching |
| Storage approach | Striping |
| | I/O parallelism |
| | Block/Fragment size |
| | Access patterns optimization |
| | Hardware optimizations |
| Network approach | Specialized network protocol |
| | Caching |
| | Multiple metadata servers |
| | Specialized network |
| | Multipathing |
| | Load balancing |

The performance optimization features are includes caching, storage and network approaches. It suggests making file data and metadata caching as a way of performance improvement of metadata operations and operations of user data access. Moreover, it makes sense to use some technique of write operations caching. A promising way of performance improvement on storage level is striping and using I/O parallelism techniques. Large block size can be a good basis for getting data and allocation of free space. It can be suggested an optimization for the most common metadata access scenarios. A hardware way of optimization is always the best way of achieving the performance breakthrough on storage level (multipathing, multiple RAID devices pool and so on). A network way of performance improvement can be seen in: (1) network protocol improvement (for example, in network protocol simplification); (2) using specialized networks (for example, dedicated metadata and data networks); (3) using multipathing (multiple servers in cluster, striping data between many servers and so on); (4) using caching as a way of read-ahead or asynchronous sending blocks to DataNodes; (5) using load balancing as a way of achieving an even balance between all nodes in a cluster.

Table 32 Reliability Features of AverageDFS

| colspan="2" | AverageDFS: Reliability Features |
|---|---|
| Hardware Approach | RAID |
| | Failover |
| Storage Approach | Snapshot |
| | Journaling |
| | Checksumming |
| Filesystem Approach | Logging |
| | Journaling |
| Network Approach | Data transfer protection |
| | Replication |

The reliability of file system can be achieved by means of Redundant Array of Independent Disks (RAID) and different failover techniques on hardware level. The storage level can be reliable on the basis of using snapshots (as point in time copy of a file system), journaling approach (as a way of, for example, saving in journal unfinished transactions), checksums (as a way of detect corruption of stored data). The level of block-oriented file system can be enhanced by using logging (as a way of saving file system's modifications in the form of log adding) or/and by means of journaling approach. Replication and data transfer protection can be used as reliability approaches on the network level.

Table 33 High-Availabilty Features of AverageDFS

| colspan="2" | AverageDFS: High-Availability Features |
|---|---|
| Network | Snapshot/Recovery |
| | Replication |
| | Load balancing |
| | Nodes monitoring |
| | Failover |
| Hardware | Multipathing |
| | Component redundancy |
| | Dedicated network |

High-availability of file system can be based on multipathing, component redundancy and dedicated network approaches on the hardware level. Dedicated network can be used for decouple different types of traffic or/and be used as management network for automatic reboot and shutdown of suspected nodes. Snapshoting with addition of automatic background recovery can be used as failover technique of transparent recovery of failed nodes in the cluster. Replication of data blocks is a policy that has goal to guarantee a data blocks availability in the environment of possible nodes' failure. Load balancing is approach of even distribution of load or data blocks between nodes with the purpose to improve cluster performance and decrease a probability of bottlenecks occurrence. Nodes monitoring is approach of active tracking nodes' state, detection of failed nodes and initiating of actions that can guarantee of cluster's high-availability. Techniques of failover are different approaches of reaction on nodes failure with the goal to guarantee of cluster high-availability by means of service migration between nodes.

Table 34 Synchronization Features of AverageDFS

| | **AverageDFS: Synchronization Features** |
|---|---|
| Synchronization techniques | Lock |
| | Lease |
| | Token mechanism |
| Application level mechanisms | Application level mechanisms |
| Notification mechanisms | Notification mechanisms |
| Transaction approach | Transaction approach |
| No lock approach | No lock approach |

Factually, this set of synchronization features (see, Table 34) can be seen as competitive approaches. But using these approaches in different environments/workloads can be mutually complimentary approaches of file system's synchronization technique.

Figure 24 AverageDFS Features' Weight Comparison

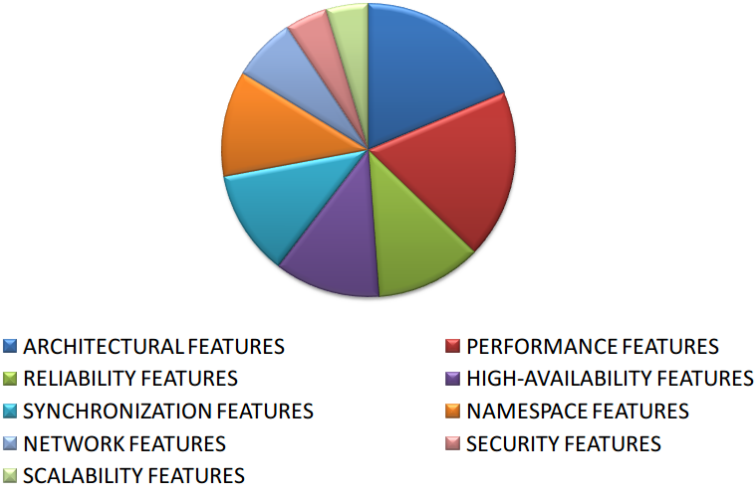

It was made also comparison of weight of features of different classes after distinction of features with the largest weight in a feature class (see, Figure 24). It is possible to see that modern DFS should have security and scalability features. A base architectural and performance optimization approaches can be a most important and time-consuming features. Namespace, reliability, high-availability and synchronization features have approximately identical importance. Moreover, a reasonable correlation of performance, namespace, reliability, high-availability и synchronization features can be achieved by means significant efforts in elaboration of good architectural and performance optimization approaches.

## 6.4 StrangeDFS

Suppose that it needs to elaborate vision of file system architecture with compilation of all original features of other file systems. Thereby, it needs to choose internal techniques with the least weight in the feature class. Such set of internal techniques will be a vision of architecture of abstract DFS (StrangeDFS).

The choosing of features with the least weight in the feature class can have a goal to distinguish a vision of some special internal technique in a feature class. Confrontation and comparative analysis of features in concrete sub-class can give opportunity to distinguish an internal technique that can solve a bottleneck or to suggest a special idea. Moreover, such comparison can give opportunity to suggest research direction with the purpose to elaborate special internal technique.

**Table 35 Architectural Features of StrangeDFS**

| StrangeDFS: Architectural Features | |
|---|---|
| Specialized API Approach | Not POSIX-compatible API |
| | Specialized Data Management API |
| Pattern Approach | Specialized Access Pattern |
| | Decouple data and metadata operations |
| Server Approach | Multiple NameNodes |
| | Object Store Devices |
| Storage Approach | RAID |
| | Striping |
| | Garbage Colection |
| Specialized network | Specialized network |

**Table 36 Performance Optimization Features of StrangeDFS**

| StrangeDFS: Performance Optimization Features | |
|---|---|
| Specialized API Approach | Specialized interfaces for data access patterns |
| Metadata Approach | Knowledge of metadata popularity |
| | Metadata load balancing |
| | Hashing to organize directory entries within a directory |
| Caching Approach | I/O parallelism |
| | Preallocation |
| | Coherent cache (cooperative cache) |
| | Lazily flushed journals strategy |
| | Metadata access scenarios optimization |
| | Transactions batching |
| Multi-threading Approach | Stateless servers and clients |
| | Idle Resource Utilization |
| Storage Approach | Specially optimized filesystem |
| | Specially optimized low-level disk scheduler |
| | Special I/O modes |
| | Load balancing |
| Network Approach | Scheduling a task to the data location |
| | Chunk size |
| | Server replication |
| | Specialized replication scheme |
| | Multiple metadata servers |
| | Specialized network |

**Table 37 Reliability Features of StrangeDFS**

| StrangeDFS: Reliability Features | |
|---|---|
| Hardware Approach | Dual access |
| | UPS notifications |
| | Nodes self-report |
| | Active monitoring |
| Storage Approach | Mirroring |
| | Transactions |
| | Replacement blocks for the file region allocation |
| | Cache entire files |
| File System Approach | Replication |
| | Checkpoint |
| | Metadata checksums |
| Network Approach | Specialized network |
| | Synchronous store operations |
| | Logging |
| | Distributed transactions |
| | Failover by means of quick rescan of the journal by another node |

| | Disconnected operations |
|---|---|

**Table 38 High-Availability Features of StrangeDFS**

| colspan="2" | StrangeDFS: High-Availability Features |
|---|---|
| Network approach | Snapshot/Recovery |
| | Load Balancing |
| | Elections |
| | Disconnected operations |
| | Garbage collection |
| Hardware approach | Components redundancy |
| | Dedicated network |

**Table 39 Synchronization Features of StrangeDFS**

| colspan="2" | StrangeDFS: Synchronization Features |
|---|---|
| Data Processing Pattern | Data Processing Pattern |
| Replica Processing | Replica Processing |
| Disconnected operations | Disconnected operations |
| Request Approach | Request Approach |